\documentclass[smallextended]{svjour3}       % onecolumn (second format)
\usepackage{graphicx}% Include figure files
\usepackage{dcolumn}% Align table columns on decimalpoint
\usepackage{bm}% bold math
\usepackage{graphicx}
\usepackage{latexsym,color}
\usepackage{amssymb}
\usepackage[colorlinks,citecolor=blue,urlcolor=blue,linkcolor=blue]{hyperref}
\usepackage{graphicx}% Include figure files
\usepackage{bm}% bold math
\usepackage{ctable}
\usepackage{amssymb}
\usepackage{amsmath}
\usepackage{hyperref}
\usepackage{latexsym,color}
\usepackage{amssymb}
\begin{document}
\newcommand{\ti}[1]{\mbox{\tiny{#1}}}
\def\be{\begin{equation}}
\def\ee{\end{equation}}
\def\bea{\begin{eqnarray}}
\def\eea{\end{eqnarray}}
\def\beas{\begin{eqnarray*}}
\def\eeas{\end{eqnarray*}}
\newcommand{\tb}[1]{\textbf{\texttt{#1}}}
\newcommand{\rtb}[1]{\textcolor[rgb]{1.00,0.00,0.00}{\tb{#1}}}
\newcommand{\btb}[1]{\textcolor[rgb]{0.00,0.00,1.00}{\tb{#1}}}
\newcommand{\il}{~}
\newcommand{\dd}{\mathcal{D}}
\def\0o{\overset{o}}
\newcommand{\lie}{\mathcal{L}}
\newcommand{\la}{\mathcal{A}}
\newcommand{\Tem}{T^{\rm{em}}}
\newcommand{\g}[1]{\Gamma^{\phantom\ #1}}
\newcommand{\para}{{}^{\ti{$\parallel$}}}
\newcommand{\ort}{{{}^{\ti{$\perp$}}}}
\newcommand{\rin}[1]{\mathring{#1}}
\newcommand{\oB}{\0o{\mathbf{B}}}

\title{Deformations of three-dimensional metrics}
\author{Daniela Pugliese \and
        Cosimo Stornaiolo}

\institute{Daniela Pugliese \at
              Institute of Physics, Faculty of Philosophy \& Science,
  Silesian University in Opava,
 Bezru\v{c}ovo n\'{a}m\v{e}st\'{i} 13, CZ-74601 Opava, Czech Republic\\
              \email{d.pugliese.physics@gmail.com }           %  \\
           \and
           Cosimo Stornaiolo \at
              INFN, Sez. di Napoli
              Complesso Universitario MSA
              via Cintia
              80126 Napoli, Italy\\
               \email{cosmo@na.infn.it}                  }%
%\date{\today}
\date{Received: date / Accepted: date}

\maketitle
\begin{abstract}
We examine  three-dimensional  metric deformations based on a tetrad transformation { through} the action the matrices of scalar field. We describe by this approach to deformation the results obtained   by Coll et al. in  \cite{three--dim},  where it is stated that any three--dimensional metric  was  locally obtained as a deformation of a constant curvature metric parameterized by a 2--form.
To this aim, we construct  the corresponding  deforming matrices and provide their classification according to the properties  of  the scalar $\sigma$ and of the vector $\mathbf{s}$ used  in  \cite{three--dim} to deform the initial metric.  The resulting  causal structure of the deformed geometries is  examined, too.
 Finally we apply our results to  a  spherically symmetric three geometry and to a  space sector of Kerr  metric.
\end{abstract}
%%%% ----------------------------------------------------------------------
%\maketitle
\section{Introduction}
Differently from other classical field theories there is not a general solution of the Einstein equations. Frequently it is necessary to impose some symmetries on space-time to find exact  solutions. Despite the restrictions introduced considering the symmetries, exact solutions are important in general relativity, as well as in extended gravity,   because they provide  a deep insight of the gravitational issues.   Many general properties of the gravitational field  would not have been well understood  without knowing the exact solutions of general relativity. Important examples are the physics of  black holes and development of the cosmological models.   In addition by  comparing the solutions of general relativity with the corresponding solutions the extended gravity theories has allowed us to understand the fundamental differences between the various theories and to formulate new scenarios in a cosmological context in addressing the dark matter and dark energy problems.

However  real space-time never shows a perfect symmetry. Deviations from symmetry may be small, but still significative. Just to give an example if we want to make good geodesy on the earth it is necessary to consider the earth as a geoid, whose deviations from being an exact ellipsoid are $\delta R/R \sim 10^{-5}$,   where $R$ is the average radius of the earth. Similarly  the deviations of the cosmic gravitational background temperature are of the order  of $\delta T/T \sim 10^{-5}$.  which means that  trying to introduce a more accurate metric is a problem that can have a certain interest in the description of the universe, especially in the era of precision cosmology. How to fit an idealized exactly homogeneous and isotropic solution of the Einstein equations to a lumpy universe was considered in  \cite{Ellis:1987zz} by examining  averaging, normal coordinates and null data and trying to relate them to data coming from astronomical observations.

Along with the exact solutions we must also consider the solutions obtained by slightly modifying the space-time background. For example gravitational waves were found by considering the propagation of small deviations from the Minkowski background. In the cosmological models. The perturbations introduced into the bottom of homogeneous and isotropic in fact lead us to consider new solutions of Einstein's equations. But these solutions are not completely covariant in the spirit of Einstein's theory as perturbations that are small in a particular coordinate system can be made arbitrarily large with particular changes of coordinates.

We can deal with these problems in general by introducing the concept of deformation of  a space-time metric, covariant way with respect to the coordinate transformations.  By deformations we mean any kind of transformation of the metric which modifies the geometrical properties of the original space-time.  In other words  a deformation is an application  defined on the superspace which associates to each metric (modulo diffeomorphisms) another metric (modulo diffeomorphisms).

Deformations have been widely considered in the literature.  Conformal transformations are a first example of metric deformations. They have the important property that they preserve the causal structure of space-time. For recent literature in the subject see for example \cite{AcenaKroon11,Frie02G,LuKroon12G,LuKroon13An,LuKroon13,CFEBook}.  {The  metric deformations so defined provide  a valuable method for the investigation of some local and global  properties of space-times. For example in the investigation of stability of space-times with stationary singularities \cite{AcenaKroon11,Frie02G,LuKroon12G,LuKroon13An,LuKroon13,CFEBook,Capozziello:2009dz,Capozziello:2011et,TerKazarian:2011kh}.} Noteworthy the solutions of the $f(R)$ theories and the corresponding solutions of Einstein equations are related by a conformal transformation.

{Another way to obtain a deformed manifold is} by considering the Ricci flow, i.e.  a flow on the space of metrics (in different space-time dimensions)  described by a parabolic equation.   It makes sense for compact manifolds and it describes the expansion of  negatively curved regions of the manifold and the contraction of positively curved regions. Imposing the appropriate corrections it can describe the deformation of a manifold to a smooth manifold. Other similar flows have been introduced in literature like the Yamabe flow and the Calabi flow.
Ricci flows has been used to discuss a smoothing procedure deforming a family of inhomogeneous cosmological models into a FLRW universe which describes an average behavior of the more realistic universe \cite{carfandmarz}.  On the other hand  perturbations in a homogeneous and isotropic universe  lead  to a deformation of the FLRW metric (see for example \cite{Mukhanov:1990me}).
A relevant example of deformation of a metric to another was given by Newman and Janis  in \cite{Newman:1965tw}, where they showed that the Kerr metric can be obtained by a simple complex transformation of the Schwarzschild metric.  The same algorithm was used to find the Kerr-Newman metric.
The construction of  the BTZ solution for a black hole in \cite{Carlip:1994gc} may be considered as a deformation of the AdS metric . Finally we mention that deformation  of a metric by introducing  torsion was studied in \cite{TerKazarian:2011kh}.

It seems necessary to give a general description of a metric deformations. A first attempt was given by the work of Coll and collaborators  \cite{three--dim} and  \cite{LlosaeSoler,Soler:2005xa,Llosa:2008yk,Llosa:2009sc,Coll:2000rm,Coll1}, where a particular definition was given to the deformations of a space-time metric.  They showed how a generic metric can be related to a metric with constant curvature and tried to extend this results to any dimension.

In  \cite{Cap-S07,Tesi}   space-time  metric deformations  were introduced  in a more general way.  Such deformations are defined in Sec.\il(\ref{Sec:sef-sca}), they are based on the introduction of matrices of scalar fields which transform  a tetrad  taking a linear combination of the vectors (or respectively convectors) at each point of the space-time. It results that  we find a new  metric  by reconstructing it with the new tetrad.
In this paper we study  the deformations of a 3-metric, as defined in \cite{Cap-S07,Tesi}, in relation with the work of   Coll and collaborators.
As a result we  find a broad classification of metric deformations and we analyze the consequences to the  casual structure of the deformed
manifold.
Our results are covariant with respect to coordinate transformations, even if  the price payed for this is  that the matrices introduced are well-defined except for a left Lorentz transformation, but such gauge freedom disappears after constructing the deformed metric.

\medskip

This article is organized as follows:
in Sec.\il(\ref{Sec:sef-sca})  we briefly introduce our definition of  space-time deformations. In Section (\ref{seza1}) we will consider
{deformations of three--dimensional metrics}, we first present a
general procedure outlined in \cite{LlosaeSoler}.
{Deforming matrices for three--dimensional metrics} are then  discussed  in Sec.\il(\ref{Sec:three-metric}), exploring  in particular the real solutions in Sec.\il(\ref{Sec:real-s}),  and the
{ complex solutions} in Sec.\il(\ref{Sec:complexone}).
In  Sec.\il(\ref{Sec:thecausalstructure}) we investigate  the casual structure of a deformed
manifold compared with the original one and finally in Sec.\il(\ref{SeC:example}) we shall consider the  deformations  of some
three--dimensional Riemannian manifolds  into flat metrics, applying
 the results  to the metric of a the three--dimensional sphere  in Sec.\il(\ref{Sec:3-sferical}) and   to the three--dimensional Kerr space in Sec.\il(\ref{Sec:Kerr}). In Section\il(\ref{Sec:conclusion-disc}) some concluding remarks are presented.
\section{Deformation of metrics}\label{Sec:sef-sca}

As stated in the introduction in order to deform a metric we look for   an application  defined on the superspace which associates to each metric (modulo diffeomorphisms) another metric (modulo diffeomorphisms).

One way to deform a metric is the following,
 let us consider the decomposition of  a  covariant metric $g$ on an n--dimensional
manifold $M$ in terms of a cotetrad field  $\omega_{a}^{\mbox{\tiny{A}}}$ and correspondingly the decomposition of the controvariant metric in terms of  the tetrad field $e^{a}_{\mbox{\tiny{A}}}$  (such that  $\omega_{a}^{\mbox{\tiny{A}}}e^{a}_{\mbox{\tiny{B}}}=\delta^{\mbox{\tiny{A}}}_{\mbox{\tiny{B}}}$)

%\mbox{\tiny{}}
\begin{equation}\label{Lep-lo}
g_{ab}=\eta_{\mbox{\tiny{AB}}}\omega_{a}^{\mbox{\tiny{A}}}\omega_{b}^{\mbox{\tiny{B}}},
\quad
g^{ab}=\eta^{\mbox{\tiny{AB}}}e^{a}_{\mbox{\tiny{A}}}e^{b}_{\mbox{\tiny{B}}},
\end{equation}
 where   $\eta_{\mbox{\tiny{AB}}}$ is the Minkowski metric in tetrad
indices\footnote{ Capital latin letters $A$ are tensorial indices while Latin letter $a$  denote the tensorial character of each object —i.e. they are spacetime indices.}.
 Then, given a  matrix $\phi^{ \mbox{\tiny{A}}}_{ \phantom\
\mbox{\tiny{B}}}$ of scalar fields defined on $M$, which is different from a Lorentz matrix at least in one point, we obtain a covariant tensor $\widetilde{g}_{ab}$
\begin{equation}\label{don-v-ra}
\widetilde{g}_{ab}\equiv \eta_{\mbox{\tiny{AB}}}\phi^{
\mbox{\tiny{A}}}_{\phantom\ \mbox{\tiny{C}}}\phi^{
\mbox{\tiny{B}}}_{\phantom\
\mbox{\tiny{D}}}\omega_{a}^{\mbox{\tiny{C}}}\omega_{b}^{\mbox{\tiny{D}}},
\quad
\end{equation}
which can be considered as deformed  metric in a different manifold $\tilde{M}$
together with the  frames defined by the deformed cotetrad field and
$\tilde{\omega}^{\mbox{\tiny{B}}}_{a}$ and tetrad field
$\tilde{e}_{\mbox{\tiny{A}}}^{a}$ defined by the
compositions
\begin{eqnarray}\label{Ze-don-Al}
\tilde{\omega}^{\mbox{\tiny{B}}}_{a}\equiv\omega^{\mbox{\tiny{A}}}_{a}\phi^{\mbox{\tiny{B}}}_{\phantom\
\mbox{\tiny{A}}}, \quad
\tilde{e}_{\mbox{\tiny{A}}}^{a}\equiv\psi^{\phantom\
\mbox{\tiny{B}}}_{ \mbox{\tiny{A}}}e_{\mbox{\tiny{B}}}^{a},
\end{eqnarray}
where $\psi^{\phantom\ \mbox{\tiny{B}}}_{ \mbox{\tiny{A}}}$ is a
second matrix of scalar fields.
%Where  the scalar fields $\phi$ and $\psi$ are not
%singular  the sets of vectors (\ref{Ze-don-Al})  form two
%basis of deformed  vectors, and tensor (\ref{don-v-ra}) is  a deformed metric; note that by
%transformation  (\ref{Mas-tt}) the deformed tensor could have a
%signature different from the original metric; for example this could
%happen with complex first deforming matrices.
We shall consider in the following that $\phi$ and $\psi$ satisfy the condition
\begin{equation}\label{don-A}
\psi^{\mbox{\tiny{A}}}_{\phantom\  \mbox{\tiny{B}}}=\phi^{-1
\mbox{\tiny{A}}}_{ \phantom\ \mbox{\tiny{B}}}.
\end{equation}
so that  the following relations hold
\begin{equation}\label{C-ra}
\tilde{\omega}^{\mbox{\tiny{A}}}_{a}\tilde{e}_{\mbox{\tiny{B}}}^{a}=\delta^{\mbox{\tiny{A}}}_{\mbox{\tiny{B}}},\quad
\tilde{\omega}^{\mbox{\tiny{A}}}_{a}\tilde{e}_{\mbox{\tiny{A}}}^{b}=\delta^{b}_{a}.
\end{equation}
%the first deforming matrices $\phi$ and $\psi$
Thus, the metric $\widetilde{g}_{ab}$ and the controvariant
tensor $\widetilde{g}^{ab}$ is  defined respectively by (\ref{don-v-ra}) as
\begin{equation}\label{Mas-tt}
\widetilde{g}_{ab}=\eta_{\mbox{\tiny{AB}}}\phi_{\phantom\
\mbox{\tiny{C}}}^{ \mbox{\tiny{A}}}\phi_{\phantom\
\mbox{\tiny{D}}}^{
\mbox{\tiny{B}}}\omega_{a}^{\mbox{\tiny{C}}}\omega_{b}^{\mbox{\tiny{D}}},\quad
\quad\widetilde{g}^{ab}=\eta^{\mbox{\tiny{AB}}}{\phi^{-1}}^{\phantom\
\mbox{\tiny{C}}}_{ \mbox{\tiny{A}}}{\phi^{-1}}^{\phantom\
\mbox{\tiny{D}}}_{
\mbox{\tiny{B}}}e^{a}_{\mbox{\tiny{C}}}e^{b}_{\mbox{\tiny{D}}} ,
\end{equation}
or
\begin{equation}\label{sve-rato}
\widetilde{g}_{ab}=\eta_{\mbox{\tiny{AB}}}
\widetilde{\omega}_{a}^{\mbox{\tiny{A}}}\widetilde{\omega}_{b}^{\mbox{\tiny{B}}}\,
, \quad\widetilde{g}^{ab}=\eta^{\mbox{\tiny{AB}}}
\tilde{e}^{a}_{\mbox{\tiny{A}}}\tilde{e}^{b}_{\mbox{\tiny{B}}}\,
\end{equation}
On the other hand the previous relations can be also read as deformations of the Minkowski metrics $\eta_{\mbox{\tiny{AB}}} $ and  $\eta^{\mbox{\tiny{AB}}} $   by   a second couple of symmetric matrices of
scalar fields, $\mathcal{G}_{\mbox{\tiny{CD}}}$ and
$\mathcal{G}^{\mbox{\tiny{CD}}}$ defined on the manifold $M$
respectively as
\begin{equation}\label{sven-ra}
\mathcal{G}_{\mbox{\tiny{CD}}}\equiv\eta_{\mbox{\tiny{AB}}}\phi_{\phantom\
\mbox{\tiny{C}}}^{ \mbox{\tiny{A}}}\phi_{\phantom\
\mbox{\tiny{D}}}^{ \mbox{\tiny{B}}},\quad\textrm{and}\quad
\mathcal{G}^{\mbox{\tiny{CD}}}\equiv\eta_{\mbox{\tiny{AB}}}\psi_{\mbox{\tiny{A}}}^{\phantom\
\mbox{\tiny{C}}}\psi^{\phantom\ \mbox{\tiny{D}}}_{\mbox{\tiny{B}}}
\end{equation}
%of the second deforming matrices
and then   writing  the metrics $\tilde{g}$    as
\begin{equation}\label{cam-pe}
\widetilde{g}_{ab}=\mathcal{G}_{\mbox{\tiny{CD}}}\omega_{a}^{\mbox{\tiny{C}}}\omega_{b}^{\mbox{\tiny{D}}},\quad
\widetilde{g}^{ab}=\mathcal{G}^{\mbox{\tiny{CD}}}e^{a}_{\mbox{\tiny{C}}}e^{b}_{\mbox{\tiny{D}}}.
\end{equation}
%
%in terms of the {\em second deforming matrices}
%$\mathcal{G}_{\mbox{\tiny{CD}}}$ and
%$\mathcal{G}^{\mbox{\tiny{CD}}}$.
We do not impose particular requirement on the matrices $\phi^{ \mbox{\tiny{A}}}_{ \phantom\
\mbox{\tiny{B}}}$ except they have to be invertible. They may even be singular in some point, for example by deforming a Minkowski space-time to a Schwarzschild space-time,  inevitably we must take a singular matrix.

Finally we note that since the Minkowski metric is invariant by Lorentz transformations, that
\begin{equation}\label{Mas-tt1}
\widetilde{g}_{ab}=\eta_{\mbox{\tiny{AB}}}\phi_{\phantom\
\mbox{\tiny{C}}}^{ \mbox{\tiny{A}}}\phi_{\phantom\
\mbox{\tiny{D}}}^{
\mbox{\tiny{B}}}\omega_{a}^{\mbox{\tiny{C}}}\omega_{b}^{\mbox{\tiny{D}}}=
\eta_{\mbox{\tiny{AB}}}
\Lambda_{\phantom\
\mbox{\tiny{D}}}^{ \mbox{\tiny{A}}}\phi_{\phantom\
\mbox{\tiny{C}}}^{ \mbox{\tiny{D}}}
\Lambda_{\phantom\
\mbox{\tiny{F}}}^{
\mbox{\tiny{B}}}\phi_{\phantom\
\mbox{\tiny{D}}}^{
\mbox{\tiny{F}}}\omega_{a}^{\mbox{\tiny{C}}}\omega_{b}^{\mbox{\tiny{D}}},\quad
\end{equation}
where the matrices $\Lambda^{\mbox{\tiny{A}}}_{\phantom\
\mbox{\tiny{B}}}$ which are Lorentz at each point of the manifold.
Then the deformation matrices are not uniquely defined and to each metric deformation we have to associate a family of matrices.

Particular simple  deformations  are the
 deformations  form $\phi_{\phantom\ \mbox{\tiny{C}}}^{
\mbox{\tiny{A}}}(x)=\Omega(x)\delta_{\phantom\ \mbox{\tiny{C}}}^{
\mbox{\tiny{A}}}$, and more generally $\phi_{\phantom\ \mbox{\tiny{C}}}^{
\mbox{\tiny{A}}}(x)=\Omega(x)\Lambda_{\phantom\ \mbox{\tiny{C}}}^{
\mbox{\tiny{A}}}$, which generate a conformal transformation of the metric with conformal factor $\Omega^{2}(x)$.

\section{Deformations of three--dimensional metrics}\label{seza1}
In the context of deformations of three--dimensional metrics Coll et al.
{\small{\cite{three--dim}}} proved  the following theorem:
\newtheorem{Theorem}{Theorem}
\begin{Theorem}\label{th1}
Let $M$ be  a three--dimensional Riemannian manifold, and $g$ a generic
3-dimensional metric. Then $g$ may be locally obtained from a
constant curvature metric $h$ by a deformation of the form
\begin{equation}\label{eqn:eulero}
g=\sigma h+\epsilon\; \textbf{s}\otimes\textbf{s}\ ,
\end{equation}
where $\sigma$ and $\mathbf{s}$ are respectively a scalar function
and a differential 1--form on $M$ and\ $\epsilon=\pm1 $.
\end{Theorem}
 Let us note  that,  fixed $g$ and $h$,  $\sigma $ and
$\mathbf{s}$ are not independent as, by Riemann's theorem,  a
generic n--dimensional metric has $ f= n(n-1)/2 $ degrees of
freedom. Thus, according to it,  a scalar relation
$\Psi(\sigma,\|\mathbf{s}\|)=0$ between $\sigma$ and $\mathbf{s}$
has to be imposed  and then the metric can be defined, at most, by
three independent functions.
On the other hand   the direction $\mathbf{s}$ is
not uniquely determined by the deformed metric $g$ so that relation
(\ref{eqn:eulero}) can be achieved in an infinite number of ways.

Moreover, \textbf{Theorem} \ref{th1} establishes also the following:
\begin{Theorem}\label{th2}
Let $(M,g)$ be   a Riemannian three--manifold, there locally exist a
function  $\phi$  and a 1--form $\mu$  such that the tensor:
\begin{equation}\label{te-en-break}
 \tilde{g}:=  \phi  g - \epsilon \textbf{$\mu$}\otimes\textbf{$\mu$}
\end{equation}
with   $ \epsilon=\pm1$  is Riemannian metric with constant
curvature; It is possible then add an arbitrary relation among the
function  $\phi$ and $|\mu|^{2}:= g^{ij}\mu_{i}\mu_{i}$.
\end{Theorem}
%
%\ End theorem ()
The inverse  metric is $\tilde{g}_{ij}$
$$
h^{rj}
\equiv\phi^{-1}\left(g^{rj}+\frac{1}{\phi-m_{0}}M^{rj}\right)\ ,
$$
where $  M_{ij} \equiv \epsilon\mu_{i}\mu_{j}$ and  $ m_{0 }\equiv
g^{ij}M_{ij} = \epsilon|\mu|^{2}$. This result follows directly from
on substituting $\tilde{g}$ with $h$, $\phi^{-1}$ with $\sigma$ and
$\phi^{-1/2}\mathbf{\mu}$ with $\mathbf{s}$.
Theorem \ref{th1}  and \ref{th2} before have been proved  in \cite{three--dim} in
the case of three--dimensional Riemannian manifold. Nevertheless  Llosa
and Soler extended the results in \cite{LlosaeSoler}  to a
semi--Riemannian $n$--dimensional manifold: having in mind that any
semi--Riemannian metric in a $n$--manifold has $n(n-1)/2$ degrees of
freedom, as many as the number of component of a differential
2--form, they proved that this metric can be obtained as a
deformation of a constant curvature metric, this deformations being
parameterized by a two form \cite{LlosaeSoler}.
It is interesting to note that the Kerr--Schild class of
metrics in general relativity could be seen as particular case of
the relation (\ref{eqn:eulero}) where $\textbf{s}$ is a null vector
and $h$ is the Minkowski metric.
 On the other hand it is  well known that, according to the Gauss
theorem, any two--dimensional semi--Riemannian metric can be always
locally mapped by a conformal transformation into  Minkowski
spacetime, then it is easy  to see that Eq.\il(\ref{eqn:eulero}) could
be used as a map from three--dimensional spaces into spaces of constant
curvature, and in this sense the result (\ref{eqn:eulero})
generalizes  the Gauss theorem in three--dimensions. Since
the conformal factor $\sigma$ can be written as a function of the
$\mathbf{s}$,  the deformation (\ref{eqn:eulero}) establishes also,
at fixed $h$, a relation between the set of metrics $g$ and that of
$\mathbf{s}$ .
However we  stress that the (\ref{eqn:eulero})  is valid
only locally (in the neighborhood of the Cauchy surface) and
moreover the proof of Theorem\il(\ref{th1}) is valid only in the
analytical case where all the data  are real analytic functions.
The law (\ref{te-en-break}), can be seen
as a deformation of a metric $g$ by a vector $\mathbf{\mu}$ to
obtain a constant metric $ \tilde{g}$, has not a unique solution
but, a part the freedom of the choice of the constraint $\Psi$, a
great arbitrariness is left also in the choice of the Cauchy data
(see \cite{LlosaeSoler}).
As  pointed out in  {{\cite{LlosaeSoler}}} and as it will
be discussed  in the following analysis, as a consequence of this
arbitrariness  there is a great variety of vectors $\mathbf{s}$
that deform a given constant curvature metric into another metric
which has the same constant curvature.
\section{Deforming matrices for three--dimensional metrics}\label{Sec:three-metric}
In this section we  focus on
{Theorem}  \ref{th1} looking for the deforming matrices
associated with the transformation (\ref{eqn:eulero}), in particular we
 find the explicit expression  for the generic element
of the deforming matrix associated to (\ref{eqn:eulero}),  distinguishing the complex scalar
fields from the real ones.
First  we remark that Theorem \ref{th1} is proved locally in the
analytical case and moreover it defines a  constant curvature  Riemannian metric.
Deformation
(\ref{Mas-tt}) relates
two metric tensors by means of generic, non necessarily real, scalar fields; this definition
is apparently universal, in the sense that it is able to be applied in any
 manifold $M$ by means of  a general choice of the  triad and the scalar fields
defined on $M$.
Hence, here  we restrict ourselves to the case of constant curvature
metric.
 In particular
we prove that, since Eq.\il(\ref{eqn:eulero}) can be recovered from
(\ref{Mas-tt}),   this can be regarded  as a particular case of
deformation (\ref{sve-rato}).
 Let us write the metrics  $h$ and  $g$ in the
(\ref{eqn:eulero}) in the triad of covectors
$\omega_{b}^{\mbox{\tiny{D}}}$, as:
\begin{eqnarray}\label{scr-va}
h_{ab}&=&\eta_{\mbox{\tiny{AB}}}
\omega_{a}^{\mbox{\tiny{A}}}\omega_{b}^{{\mbox{\tiny{B}}}},\quad
g_{ab}=\mathcal{G}_{\mbox{\tiny{CD}}}
\omega_{a}^{\mbox{\tiny{C}}}\omega_{b}^{\mbox{\tiny{D}}}
\end{eqnarray}
where $\mathcal{G}_{\mbox{\tiny{CD}}}$ is a second deforming matrix.
Equation (\ref{scr-va}) can be written in terms of the   deforming
matrix\ $\phi_{\phantom\ \mbox{\tiny{C}}}^{\mbox{\tiny{A}}}$ as
follows
\begin{equation}\label{spa-rti}
g_{ab}=\eta_{\mbox{\tiny{AB}}} \phi_{\phantom\
\mbox{\tiny{C}}}^{\mbox{\tiny{A}}}\phi_{\phantom\
\mbox{\tiny{D}}}^{\mbox{\tiny{B}}}\omega_{a}^{\mbox{\tiny{C}}}\omega_{b}^{\mbox{\tiny{D}}}
\end{equation}
%Chiedi
Thus, comparing Eq.(\ref{spa-rti}) with the
Eq.(\ref{eqn:eulero}), we infer that the following identity should always
hold
\begin{equation} \label{eq}
g_{ab}=\sigma h_{ab}+\epsilon s_{a}s_{b}=\eta_{\mbox{\tiny{AB}}}
\phi_{\phantom\ \mbox{\tiny{C}}}^{\mbox{\tiny{A}}}\phi_{\phantom\
\mbox{\tiny{D}}}^{\mbox{\tiny{B}}}\omega_{a}^{\mbox{\tiny{C}}}\omega_{b}^{\mbox{\tiny{D}}}.
\end{equation}
In order to solve Eq.(\ref{eq}) respect to the  variable
$\phi_{\phantom\ \mbox{\tiny{C}}}^{\mbox{\tiny{A}}}$, we
consider  the fields  $\sigma$ and   $\mathbf{s}$ as known terms;
thus the solution will  give  the field $\phi_{\phantom\
\mbox{\tiny{C}}}^{\mbox{\tiny{A}}}$ in (\ref{spa-rti}) as a
function of  $\sigma$ and $\mathbf{s}$. It is important to note that
in the following analysis we don't fix the gauge condition as  a
scalar relation $\Psi(\sigma,\|\mathbf{s}\|)=0$ between $\sigma$ and
$\mathbf{s}$; nevertheless,  it should be stressed that the
deforming matrices  are not uniquely defined but, since they form a
right coset for right composition with Lorentz matrices
$\Lambda_{\phantom\ \mbox{\tiny{C}}}^{\mbox{\tiny{A}}}$, every
element we are going to obtain from the Eq.(\ref{eq}) identifies all
the matrices of its equivalent class.
We therefore  consider a part the case
of  real solutions in Sec.\il(\ref{Sec:real-s}), and the complex scalar field case in Sec.\il(\ref{Sec:complexone}).
%
%\begin{description}
\subsection{Real solutions}\label{Sec:real-s}
%\item[1$^{\circ}$ case:] \textbf{Real solutions.}
In this first case we study the real solutions of
Eq.(\ref{spa-rti}).
In order to do this, we consider the following expression for
the real scalar field  $\phi_{\phantom\
\mbox{\tiny{C}}}^{\mbox{\tiny{A}}}$:
\begin{equation}\label{per-ta}
\phi_{\phantom\
\mbox{\tiny{C}}}^{\mbox{\tiny{A}}}=\sqrt{\sigma}\delta_{\phantom\
\mbox{\tiny{C}}}^{\mbox{\tiny{A}}}+\alpha
s^{\mbox{\tiny{A}}}s_{\mbox{\tiny{C}}} \, ,
\end{equation}
%
%$\Rset$
where  $ \sigma>0$  and   $\alpha $ is a scalar field. Therefore the
second deforming matrix is:
\begin{eqnarray}
\mathcal{G}_{\mbox{\tiny{CD}}}&=&\eta_{\mbox{\tiny{AB}}}(\sqrt{\sigma}\delta_{\phantom\
\mbox{\tiny{C}}}^{\mbox{\tiny{A}}}+\alpha
s^{\mbox{\tiny{A}}}s_{\mbox{\tiny{C}}})
(\sqrt{\sigma}\delta_{\phantom\
\mbox{\tiny{D}}}^{\mbox{\tiny{B}}}+\alpha
s^{\mbox{\tiny{B}}}s_{\mbox{\tiny{D}}}),
\end{eqnarray}
thus, we obtain
\begin{eqnarray}\label{j-E}
\mathcal{G}_{\mbox{\tiny{CD}}}&=&\eta_{\mbox{\tiny{AB}}}\sqrt{\sigma}\sqrt{\sigma}\delta_{\phantom\
\mbox{\tiny{C}}}^{\mbox{\tiny{A}}}\delta_{\phantom\
\mbox{\tiny{D}}}^{\mbox{\tiny{B}}}+
\eta_{\mbox{\tiny{}}AB}\sqrt{\sigma}\delta_{\phantom\ \mbox{\tiny{C}}}^{\mbox{\tiny{A}}}\alpha s^{\mbox{\tiny{B}}}s_{\mbox{\tiny{D}}}+\\
\nonumber &+&\eta_{\mbox{\tiny{AB}}}\alpha
s^{\mbox{\tiny{A}}}s_{\mbox{\tiny{C}}}\sqrt{\sigma}\delta_{\phantom\
\mbox{\tiny{D}}}^{\mbox{\tiny{B}}}+\eta_{\mbox{\tiny{AB}}}\alpha^{2}
s^{\mbox{\tiny{A}}}s_{\mbox{\tiny{C}}}s^{\mbox{\tiny{B}}}s_{\mbox{\tiny{D}}}\,
.
\end{eqnarray}
or also
\begin{equation}\label{ar-bi}
\mathcal{G}_{\mbox{\tiny{CD}}}=\sigma
\eta_{\mbox{\tiny{CD}}}+2\alpha\sqrt{\sigma}s_{\mbox{\tiny{C}}}s_{\mbox{\tiny{D}}}+\alpha^{2}\eta_{\mbox{\tiny{AB}}}s^{\mbox{\tiny{A}}}s^{\mbox{\tiny{B}}}s_{\ti{D}}s_{\ti{C}}\,
.
\end{equation}
Collecting the terms in (\ref{ar-bi}), we have:
\begin{equation}\label{citt-y}
\mathcal{G}_{\mbox{\tiny{CD}}}=\sigma\eta_{\mbox{\tiny{CD}}}+[2\alpha\sqrt{\sigma}+\alpha^{2}(\eta_{\mbox{\tiny{AB}}}s^{\mbox{\tiny{A}}}s^{\mbox{\tiny{B}}})]
s_{\mbox{\tiny{C}}}s_{\mbox{\tiny{D}}}.
\end{equation}
Finally, comparing Eq.\il(\ref{citt-y}) with  Eq.(\ref{eq}), we obtain the
following equation:
\begin{equation}\label{cor-re}
[2\alpha\sqrt{\sigma}+\alpha^{2}
(\eta_{\mbox{\tiny{AB}}}s^{\mbox{\tiny{A}}}s^{\mbox{\tiny{B}}})-\epsilon]
s_{\mbox{\tiny{C}}}s_{\mbox{\tiny{D}}}=0\,
,
\end{equation}
%
%\mbox{\tiny{}}
consequently
\begin{equation}\label{Curbastro}
2\alpha\sqrt{\sigma}+\alpha^{2}(\eta_{\mbox{\tiny{AB}}}s^{\mbox{\tiny{A}}}s^{\mbox{\tiny{B}}})-\epsilon=0.
\end{equation}
By setting
$\|\mathbf{s}\|^{2}\equiv\eta_{\mbox{\tiny{AB}}}s^{\mbox{\tiny{A}}}s^{\mbox{\tiny{B}}}$,
Eq.(\ref{Curbastro}) reduces to:
\begin{equation}\label{al-tro-de}
\|\mathbf{s}\|^{2}\alpha^{2}+2\sqrt{\sigma}\alpha-\epsilon=0\, .
\end{equation}
We can read Eq.(\ref{al-tro-de}) as a second order equation respect
to the variable $\alpha$, with  $\epsilon=\pm1$.

Therefore, we have the two following possibilities:
\begin{eqnarray}
\label{son-te}1.
\quad\|\mathbf{s}\|^{2}\alpha^{2}+2\sqrt{\sigma}\alpha-1&=&0\quad\textrm{for}\quad\epsilon=+1\\\nonumber\\
\label{son-te2}2.
\quad\|\mathbf{s}\|^{2}\alpha^{2}+2\sqrt{\sigma}\alpha+1&=&0\quad\textrm{for}\quad\epsilon=-1
\end{eqnarray}
In the first case the discriminant  $\Delta/4$  of the equation
(\ref{son-te}) is:
\begin{equation}\label{WWA.M}
\frac{\Delta}{4}=\sigma+\|\mathbf{s}\|^{2}.
\end{equation}
%
%%
%\begin{eqnarray}
%
%\end{eqnarray}
%%
Then we discuss  the case  ${\Delta}/{4}>0$ in Sec.\il(\ref{Sec:cad-f}) and  solutions for ${\Delta}/{4}=0$ in  Sec.\il(\ref{Sec:cad-f0}).
%
%\begin{description}
%\item[$\frac{\Delta}{4}>0$]
\subsubsection{Case: ${\Delta}/{4}>0$}\label{Sec:cad-f}
 This  first case occurs
when:
$$
\|\mathbf{s}\|^{2}>-\sigma , \quad (\epsilon=+1),
$$
In particular, if $ \|\mathbf{s}\|^{2}=0 $ the  quantity $\Delta/4$
(\ref{WWA.M}) is strictly positive, where $\sigma\neq0$ and the
solution of the first order equation Eq.(\ref{al-tro-de})  is:
$
\alpha={1}/{2\sqrt{\sigma}}
$,
and $\phi_{\phantom\ \mbox{\tiny{C}}}^{\mbox{\tiny{A}}}$ is in this
case:
\begin{equation}\label{il-fla-ma1}
\phi_{\phantom\
\mbox{\tiny{C}}}^{\mbox{\tiny{A}}}=\sqrt{\sigma}\delta_{\phantom\
\mbox{\tiny{C}}}^{\mbox{\tiny{A}}}+\frac{1}{2\sqrt{\sigma}}
\mathbf{s}^{\mbox{\tiny{A}}}\mathbf{s}_{\mbox{\tiny{C}}}.
\end{equation}
On the other hand for $\|\mathbf{s}\|\neq0$ we have two real
solutions of Eq.(\ref{son-te}), $\alpha_{\mp}$
respectively:
\begin{equation}\label{imp-sp}
\alpha_{\mp}=\frac{-\sqrt{\sigma}\mp\sqrt{\sigma+\|\mathbf{s}\|^{2}}}{\|\mathbf{s}\|^{2}};%$%\qquad%\alpha_{+}=\frac{-\sqrt{\sigma}+
%\sqrt{\sigma+\|\mathbf{s}\|%^{2}}}{\|\mathbf{s}\|^{2}}
\end{equation}
therefore the first deforming matrix $\phi_{\phantom\
\mbox{\tiny{C}}}^{\mbox{\tiny{A}}}$ is
\begin{equation}\label{fl-mag}
\phi_{\phantom\
\mbox{\tiny{C}}}^{\mbox{\tiny{A}}}=\sqrt{\sigma}\delta_{\phantom\
\mbox{\tiny{C}}}^{\mbox{\tiny{A}}}+\frac{-\sqrt{\sigma}\mp\sqrt{\sigma+\|\mathbf{s}\|^{2}}}{\|\mathbf{s}\|^{2}}
\mathbf{s}^{\mbox{\tiny{A}}}\mathbf{s}_{\mbox{\tiny{C}}}.
\end{equation}
\subsubsection{Case: ${\Delta}/{4}=0 $}\label{Sec:cad-f0}
%}
%\item[$\frac{\Delta}{4}=0 $]
This case holds when
$
 \|\mathbf{s}\|^{2}=-\sigma ,
$
therefore  the solution  of Eq.(\ref{son-te}) is
%
%\begin{equation}\label{presente}
$\alpha={1}/{\sqrt{\sigma}}.$
%\end{equation}
%
In this case the   deforming matrix is :
\begin{equation}\label{fl-mag2}
\phi_{\phantom\
\mbox{\tiny{C}}}^{\mbox{\tiny{A}}}=\sqrt{\sigma}\delta_{\phantom\
\mbox{\tiny{C}}}^{\mbox{\tiny{A}}}+\frac{1}{\sqrt{\sigma}}
\mathbf{s}^{\mbox{\tiny{A}}}\mathbf{s}_{\mbox{\tiny{C}}}.
\end{equation}
%
%\end{description}
%
Finally,  the discriminant of Eq.(\ref{son-te2})
$$
\|\mathbf{s}\|^{2}\alpha^{2}+2\sqrt{\sigma}\alpha+1=0 \quad
(\epsilon=-1)
$$
is
\begin{equation}\label{abb-ta}
\frac{\Delta}{4}=\sigma-\|\mathbf{s}\|^{2}\, .
\end{equation}
%Two particular subcases occur:
Therefore, two  subcases occur for ${\Delta}/{4}>0$ in Sec.\il(\ref{Sec:secondM}) and
 ${\Delta}/{4}=0$ in Sec.\il(\ref{Sec:secondM0}).
%
%\begin{description}
%\item[ $\frac{\Delta}{4}>0$]\
\subsubsection{Case: ${\Delta}/{4}>0$}\label{Sec:secondM}
This case holds for
$
\|\mathbf{s}\|^{2}<\sigma \quad (\epsilon=-1).$
%\end{equation}
%
% $ 0<\|\mathbf{s}\|^{2}<\sigma\,\quad \mbox{or}\quad \|\mathbf{s}\|^{2}<0\, . $
In particular, if $ \|\mathbf{s}\|^{2}=0 $ the  discriminant
(\ref{abb-ta}) is strictly positive, and the solution of the
first order equation Eq.(\ref{son-te2})  is:
$$
\alpha=-\frac{1}{2\sqrt{\sigma}}
$$
and $\phi_{\phantom\ \mbox{\tiny{C}}}^{\mbox{\tiny{A}}}$ is:
\begin{equation}\label{il-fla-ma10}
\phi_{\phantom\
\mbox{\tiny{C}}}^{\mbox{\tiny{A}}}=\sqrt{\sigma}\delta_{\phantom\
\mbox{\tiny{C}}}^{\mbox{\tiny{A}}}-\frac{1}{2\sqrt{\sigma}}
\mathbf{s}^{\mbox{\tiny{A}}}\mathbf{s}_{\mbox{\tiny{C}}}.
\end{equation}
On the other hand for $\|\mathbf{s}\|\neq0$  there are two real
solutions of Eq.(\ref{son-te2}), $\alpha_{\mp}$:
\begin{equation}\label{trascor-re}
\alpha_{\mp}=\frac{-\sqrt{\sigma}\mp\sqrt{\sigma-\|\mathbf{s}\|^{2}}}{\|\mathbf{s}\|^{2}}\ .
%,
%\qquad\alpha_{+}=\frac{-\sqrt{\sigma}+\sqrt{\sigma-\|\mathbf{s}\|^{2}}}{\|\mathbf{s}\|^{2}}\,
%.
\end{equation}
The   deforming matrix in this case is:
\begin{equation}\label{fl-mag3}
\phi_{\phantom\
\mbox{\tiny{C}}}^{\mbox{\tiny{A}}}=\sqrt{\sigma}\delta_{\phantom\
\mbox{\tiny{C}}}^{\mbox{\tiny{A}}}+\frac{-\sqrt{\sigma}\mp\sqrt{\sigma-\|\mathbf{s}\|^{2}}}{\|\mathbf{s}\|^{2}}
\mathbf{s}^{\mbox{\tiny{A}}}\mathbf{s}_{\mbox{\tiny{C}}}.
\end{equation}
%
%$$\|\mathbf{s}\|^{2}\in]-\infty,\sigma[ $\\
\subsubsection{$Case:{\Delta}/{4}=0$}\label{Sec:secondM0}
%\item[$\frac{\Delta}{4}=0$]
This case holds if
\begin{equation}\label{An-d}
\|\mathbf{s}\|^{2}=\sigma.
\end{equation}
%
%Solution of the quadratic equation yields\begin{equation}\label{costruito}
%We use the sine formula, to obtain
Thus, we find the following solution:
%Thus comparing Eq. (8) with the ã.g00 term in Eq. (3), we infer the following identities should
%always hold
\begin{equation}\label{Mon-lli}
\alpha=-\frac{1}{\sqrt{\sigma}}.
\end{equation}
in this case the  deforming matrix is:
\begin{equation}\label{fl-mag4}
\phi_{\phantom\
\mbox{\tiny{C}}}^{\mbox{\tiny{A}}}=\sqrt{\sigma}\delta_{\phantom\
\mbox{\tiny{C}}}^{\mbox{\tiny{A}}}-\frac{1}{\sqrt{\sigma}}
\mathbf{s}^{\mbox{\tiny{A}}}\mathbf{s}_{\mbox{\tiny{C}}}.
\end{equation}
%
%\end{description}
Finally, as as shown in Table\il\ref{tab:minute}, the solutions
for $\epsilon=\pm1$ exist together  for $\|\mathbf{s}\| \in
\left[-\sigma, \sigma\right]$. On the contrary,  to
$\|\mathbf{s}\|< -\sigma$    and to $\|\mathbf{s}\|>\sigma$ correspond   the solutions
$\epsilon=\mp1$  respectively.
\begin{table}[hbpt]
\begin{center}
\begin{tabular}{ccc}
\toprule
&$\epsilon=-1$ &$\epsilon=+1 $
\\
\midrule
\strut \bfseries$\|\mathbf{s}\|^{2}<-\sigma$
&$\phi_{\phantom\
\mbox{\tiny{C}}}^{\mbox{\tiny{A}}}=\sqrt{\sigma}\delta_{\phantom\
\mbox{\tiny{C}}}^{\mbox{\tiny{A}}}+\frac{-\sqrt{\sigma}\mp\sqrt{\sigma-\|\mathbf{s}\|^{2}}}{\|\mathbf{s}\|^{2}}
s^{\mbox{\tiny{A}}}s_{\mbox{\tiny{C}}} $&$\ \checkmark $
\\
\\
\strut\bfseries$\|\mathbf{s}\|^{2}=-\sigma$ &$\phi_{\phantom\
\mbox{\tiny{C}}}^{\mbox{\tiny{A}}}=\sqrt{\sigma}\delta_{\phantom\
\mbox{\tiny{C}}}^{\mbox{\tiny{A}}}+\frac{1\pm\sqrt{2}}{\sqrt{\sigma}}
s^{\mbox{\tiny{A}}}s_{\mbox{\tiny{C}}}$&$\phi_{\phantom\
\mbox{\tiny{C}}}^{\mbox{\tiny{A}}}=\sqrt{\sigma}\delta_{\phantom\
\mbox{\tiny{C}}}^{\mbox{\tiny{A}}}+\frac{1}{\sqrt{\sigma}}
s^{\mbox{\tiny{A}}}s_{\mbox{\tiny{C}}} $
\\
\\
\bfseries$\|\mathbf{s}\|^{2}\in]-\sigma,\sigma[$ &$\phi_{\phantom\
\mbox{\tiny{C}}}^{\mbox{\tiny{A}}}=\sqrt{\sigma}\delta_{\phantom\
\mbox{\tiny{C}}}^{\mbox{\tiny{A}}}+\frac{-\sqrt{\sigma}\mp\sqrt{\sigma-\|\mathbf{s}\|^{2}}}{\|\mathbf{s}\|^{2}}
s^{\mbox{\tiny{A}}}s_{\mbox{\tiny{C}}} $&$\phi_{\phantom\
\mbox{\tiny{C}}}^{\mbox{\tiny{A}}}=\sqrt{\sigma}\delta_{\phantom\
\mbox{\tiny{C}}}^{\mbox{\tiny{A}}}+\frac{-\sqrt{\sigma}\mp\sqrt{\sigma+\|\mathbf{s}\|^{2}}}{\|\mathbf{s}\|^{2}}
s^{\mbox{\tiny{A}}}s_{\mbox{\tiny{C}}} $
\\
\\
\bfseries$\|\mathbf{s}\|^{2}=0$ &$\phi_{\phantom\
\mbox{\tiny{C}}}^{\mbox{\tiny{A}}}= \sqrt{\sigma}\delta_{\phantom\
\mbox{\tiny{C}}}^{\mbox{\tiny{A}}}-\frac{1}{2\sqrt{\sigma}}s^{\mbox{\tiny{A}}}s_{\mbox{\tiny{C}}}$&$\phi^{\mbox{\tiny{A}}}_{\phantom\
\mbox{\tiny{C}}}=\sqrt{\sigma}\delta_{\phantom\
\mbox{\tiny{C}}}^{\mbox{\tiny{A}}}+\frac{1}{2\sqrt{\sigma}}s^{\mbox{\tiny{A}}}s_{\mbox{\tiny{C}}}
$
\\
\\
$\|\mathbf{s}\|^{2}=\sigma$ &$\phi_{\phantom\
\mbox{\tiny{C}}}^{\mbox{\tiny{A}}}=\sqrt{\sigma}\delta_{\phantom\
\mbox{\tiny{C}}}^{\mbox{\tiny{A}}}-\frac{1}{\sqrt{\sigma}}
s^{\mbox{\tiny{A}}}s_{\mbox{\tiny{C}}}$&$ \phi_{\phantom\
\mbox{\tiny{C}}}^{\mbox{\tiny{A}}}=\sqrt{\sigma}\delta_{\phantom\
\mbox{\tiny{C}}}^{\mbox{\tiny{A}}}+\frac{-1\pm\sqrt{2}}{\sqrt{\sigma}}
s^{\mbox{\tiny{A}}}s_{\mbox{\tiny{C}}}$
\\
\\
\strut\bfseries$\|\mathbf{s}\|^{2}>\sigma$ &$\checkmark
$&$\phi_{\phantom\
\mbox{\tiny{C}}}^{\mbox{\tiny{A}}}=\sqrt{\sigma}\delta_{\phantom\
\mbox{\tiny{C}}}^{\mbox{\tiny{A}}}+\frac{-\sqrt{\sigma}\mp\sqrt{\sigma+\|\mathbf{s}\|^{2}}}{\|\mathbf{s}\|^{2}}
s^{\mbox{\tiny{A}}}s_{\mbox{\tiny{C}}} $
\\
\\ \bottomrule
\end{tabular}
\end{center}
\caption{Deforming  matrices for three--dimensional metrics.
$1^{\circ}$ case: Real solutions.}\label{tab:minute}
\end{table}
%
%\item[2$^{\circ}$ case:] \textbf{Complex solutions}
\subsection{Complex solutions}\label{Sec:complexone}
\newcommand{\re}{\mathop{\mathrm{Re}}}
\newcommand{\im}{\mathop{\mathrm{Im}}}
In this  section we study the complex solutions of
Eq. (\ref{spa-rti}).
%\( z + \bar{z} = 2 \re z, \quad z - \bar{z} = 2 \imath \im z \)
Thus, we write  the scalar field
$\phi^{\mbox{\tiny{A}}}_{\phantom\ \mbox{\tiny{C}}}$ as:
\begin{equation}\label{P-rp}
\phi_{\phantom\
\mbox{\tiny{C}}}^{\mbox{\tiny{A}}}=\sqrt{\sigma}\delta_{\phantom\
\mbox{\tiny{C}}}^{\mbox{\tiny{A}}}+\alpha
s^{\mbox{\tiny{}}A}s_{\mbox{\tiny{C}}} \, ,
\end{equation}
where now $\alpha$ and $\sqrt{\sigma }$ are two complex scalar
fields.
Therefore in (\ref{P-rp}) we split the real part of scalar fields
from the imaginary one. We can write:
\begin{equation}\label{gr-gio}
\phi_{\phantom\
\mbox{\tiny{C}}}^{\mbox{\tiny{A}}}=(\re\sqrt{\sigma}+\imath
\im\sqrt{\sigma})\delta_{\phantom\
\mbox{\tiny{C}}}^{\mbox{\tiny{A}}}+(\re\alpha+\imath\im\alpha)
s^{\mbox{\tiny{A}}}s_{\mbox{\tiny{C}}}\, ,
\end{equation}
%\mbox{\tiny{}}
or also:
\begin{equation}\label{plo-ne}
\phi_{\phantom\
\mbox{\tiny{C}}}^{\mbox{\tiny{A}}}=\left(\re\sqrt{\sigma}\right)\delta_{\phantom\
\mbox{\tiny{C}}}^{\mbox{\tiny{A}}}+\left(\re\alpha\right) \
s^{\mbox{\tiny{A}}}s_{\mbox{\tiny{C}}}+\imath\left(
\im\sqrt{\sigma}\delta_{\phantom\
\mbox{\tiny{C}}}^{\mbox{\tiny{A}}}+\im\alpha
s^{\mbox{\tiny{A}}}s_{\mbox{\tiny{C}}}\right).
\end{equation}
We consider the case  $\sigma>0$ in Sec.\il(\ref{Sec:sM0}) and $\sigma<0$ in Sec.\il(\ref{Sec:sn0})
%and, by the explicit form of Cdab, this equation becomes
%\begin{description}
\subsubsection{Case: $\sigma>0$}\label{Sec:sM0}
%\item[$\sigma>0$]
In this first case $\sqrt{\sigma}\equiv\re\sqrt{\sigma}$  and the
expression(\ref{plo-ne}) becomes:
\begin{equation}\label{plo-ne1}
\phi_{\phantom\
\mbox{\tiny{C}}}^{\mbox{\tiny{A}}}=\sqrt{\sigma}\delta_{\phantom\
\mbox{\tiny{C}}}^{\mbox{\tiny{A}}}+\re\alpha \
s^{\mbox{\tiny{A}}}s_{\mbox{\tiny{C}}}+ {i}\im\alpha
s^{\mbox{\tiny{A}}}s_{\mbox{\tiny{C}}}\, ,
\end{equation}
therefore the second deforming matrix
$\mathcal{G}_{\mbox{\tiny{CD}}}$ is:
\begin{eqnarray}
\nonumber\mathcal{G}_{\mbox{\tiny{CD}}}&=&\eta_{\mbox{\tiny{AB}}}\left(\sqrt{\sigma}\delta_{\phantom\
\mbox{\tiny{C}}}^{\mbox{\tiny{A}}}+\re\alpha \
s^{\mbox{\tiny{A}}}s_{\mbox{\tiny{C}}}+ \imath\im\alpha
s^{\mbox{\tiny{A}}}s_{\mbox{\tiny{C}}}\right)\times
\\\label{am-ts}
&&\left( \sqrt{\sigma}\delta_{\phantom\
\mbox{\tiny{D}}}^{\mbox{\tiny{B}}}\right.+\left.\re\alpha \
s^{\mbox{\tiny{B}}}s_{\mbox{\tiny{D}}}+\imath\im\alpha
s^{\mbox{\tiny{B}}}s_{\mbox{\tiny{D}}}\right).
\end{eqnarray}
In particular for $\|\mathbf{s}\|=0$  we obtain, form
Eq.(\ref{am-ts}) the following real solutions $\phi_{\phantom\
\mbox{\tiny{C}}}^{\mbox{\tiny{A}}}$:
\begin{equation}\label{Li-qui}
\phi_{\phantom\
\mbox{\tiny{C}}}^{\mbox{\tiny{A}}}=\sqrt{\sigma}\delta_{\phantom\
\mbox{\tiny{C}}}^{\mbox{\tiny{A}}}+\epsilon\frac{1}{2\sqrt{\sigma}}s^{\mbox{\tiny{A}}}s_{\mbox{\tiny{C}}}
\end{equation}
 Finally, comparing Eq.(\ref{am-ts}) with  Eq.(\ref{eq}) we
obtain the following expressions for  $\alpha$ in the case
$\|\mathbf{s}\|\neq0$:
\begin{align}\label{rl}
\alpha_{\mp}=-\frac{\sqrt{\sigma}}{\|\mathbf{s}\|^{2}}\mp\imath\frac{1}{\|\mathbf{s}\|^{2}}
\sqrt{-(\sigma+\epsilon\|\mathbf{s}\|^{2})} %\nonumber\
%\re\alpha=-\frac{\sqrt{\sigma}}{\|\mathbf{s}\|^{2}}&\quad\textrm{and}\quad
%\im\alpha=\mp\frac{1}{\|\mathbf{s}\|^{2}}
%\sqrt{-(\sigma+\epsilon\|\mathbf{s}\|^{2})}\\
\quad\textrm{for}\quad \epsilon\|\mathbf{s}\|^{2}\leq\sigma,
\end{align}
or,
\begin{subequations}
\begin{eqnarray}
\label{titarc1}\nonumber\alpha_{\mp}&=&-\frac{\sqrt{\sigma}}{\|\mathbf{s}\|^{2}}\mp\frac{\imath}{\|\mathbf{s}\|^{2}}\sqrt{-(\sigma-\|\mathbf{s}\|^{2})},\quad
\textrm{for}\quad \epsilon=-1\quad\textrm{and}\quad
\|\mathbf{s}\|^{2}\geq\sigma.\\
\\\nonumber\\ \nonumber
\label{titarc2}\alpha_{\mp}&=&-\frac{\sqrt{\sigma}}{\|\mathbf{s}\|^{2}}\mp\frac{\mathrm
{\imath}}{\|\mathbf{s}\|^{2}}\sqrt{-(\sigma+\|\mathbf{s}\|^{2})},\quad
\textrm{for}\quad
\epsilon=+1\quad\textrm{and}\quad\|\mathbf{s}\|^{2}\leq-\sigma.\\
\end{eqnarray}
\end{subequations}
Using the Eq.\il(\ref{titarc1})   in (\ref{plo-ne1}),
we obtain  for the field $\phi_{\phantom\
\mbox{\tiny{C}}}^{\mbox{\tiny{A}}}$
\begin{align}
\label{G-NI-CRES}
\phi_{\phantom\
\mbox{\tiny{C}}}^{\mbox{\tiny{A}}}&=\sqrt{\sigma}\delta_{\phantom\
\mbox{\tiny{C}}}^{\mbox{\tiny{A}}}-\frac{\sqrt{\sigma}}{\|\mathbf{s}\|^{2}}
s^{\mbox{\tiny{A}}}s_{\mbox{\tiny{C}}}\mp\frac{\mathrm
{\imath}}{\|\mathbf{s}\|^{2}}\sqrt{-(\sigma-\|\mathbf{s}\|^{2})}
s^{\mbox{\tiny{A}}}s_{\mbox{\tiny{C}}}\,\\ \nonumber
&\textrm{for}\quad \epsilon=-1,\quad\textrm{and}\quad
\|\mathbf{s}\|^{2}\geq\sigma.
\end{align}
In particular we find the solution
\begin{align}\label{pianpta}
\phi_{\phantom\
\mbox{\tiny{C}}}^{\mbox{\tiny{A}}}&=\sqrt{\sigma}\delta_{\phantom\
\mbox{\tiny{C}}}^{\mbox{\tiny{A}}}-\frac{1}{\sqrt{\sigma}}
s^{\mbox{\tiny{A}}}s_{\mbox{\tiny{C}}}\qquad\textrm{for}\quad
\epsilon=-1,\quad \|\mathbf{s}\|^{2}=\sigma.
\end{align}
While, using the expression (\ref{titarc2})   in
(\ref{plo-ne1}), we obtain  for the field $\phi_{\phantom\
\mbox{\tiny{C}}}^{\mbox{\tiny{A}}}$:
\begin{align}\label{Ac-St-Cec}
\phi_{\phantom\
\mbox{\tiny{C}}}^{\mbox{\tiny{A}}}&=\sqrt{\sigma}\delta_{\phantom\
\mbox{\tiny{C}}}^{\mbox{\tiny{A}}}-\frac{\sqrt{\sigma}}{\|\mathbf{s}\|^{2}}
s^{\mbox{\tiny{A}}}s_{\mbox{\tiny{C}}}\mp\frac{\mathrm
{\imath}}{\|\mathbf{s}\|^{2}}\sqrt{-(\sigma+\|\mathbf{s}\|^{2})}
s^{\mbox{\tiny{A}}}s_{\mbox{\tiny{C}}}\ ,
\nonumber\\
& \textrm{for}\quad
\epsilon=+1\quad\textrm{and}\quad\|\mathbf{s}\|^{2}\leq-\sigma,
\end{align}
where in particular
\begin{align}\label{100-o}
\phi_{\phantom\
\mbox{\tiny{C}}}^{\mbox{\tiny{A}}}&=\sqrt{\sigma}\delta_{\phantom\
\mbox{\tiny{C}}}^{\mbox{\tiny{A}}}+\frac{1}{\sqrt{\sigma}}
s^{\mbox{\tiny{A}}}s_{\mbox{\tiny{C}}}\qquad \textrm{for}\quad
\epsilon=+1,\quad \|\mathbf{s}\|^{2}=-\sigma\ .
\end{align}
%\imath
\subsubsection{Case: $\sigma<0$}\label{Sec:sn0}
%\item[$\sigma<0$]
In this second case we write the second deforming matrix
$\mathcal{G}_{\mbox{\tiny{CD}}}$ as:
\begin{eqnarray}\label{cioc-to-la}
\mathcal{G}_{\mbox{\tiny{CD}}}&=&\eta_{\mbox{\tiny{AB}}}
\left[\re\sqrt{\sigma}\delta_{\phantom\
\mbox{\tiny{C}}}^{\mbox{\tiny{A}}}+\re\alpha \
s^{\mbox{\tiny{A}}}s_{\mbox{\tiny{C}}}+\mathrm {\imath}\left(
\im\sqrt{\sigma}\delta_{\phantom\
\mbox{\tiny{C}}}^{\mbox{\tiny{A}}}+\im\alpha
s^{\mbox{\tiny{A}}}s_{\mbox{\tiny{C}}}\right)\right]\\
&&\left[\re\sqrt{\sigma}\delta_{\phantom\
\mbox{\tiny{D}}}^{\mbox{\tiny{B}}}+\re\alpha \
s^{\mbox{\tiny{B}}}s_{\mbox{\tiny{D}}}+\right.\left.\mathrm
{\imath}\left( \im\sqrt{\sigma}\delta_{\phantom\
\mbox{\tiny{D}}}^{\mbox{\tiny{B}}}+\im\alpha
s^{\mbox{\tiny{B}}}s_{\mbox{\tiny{D}}}\right)\right]\ .
\end{eqnarray}
By using  Eq.\il(\ref{cioc-to-la}) in the (\ref{scr-va}) and comparing with
(\ref{eq}) we find the following two possibilities.

For  $ \|\mathbf{s}\|^{2}=0$ we find
\begin{align}
\alpha&=-\imath\,\frac{\epsilon\quad}{2
\im\sqrt{\sigma}}\quad\textrm{and}\quad\re\sqrt{\sigma}=0.
%\intertext{where}(\im\sqrt{\sigma})^{2}&=- \sigma\\
\end{align}
The deforming matrix is in this case
\begin{equation}\label{coll-lonk}
\phi_{\phantom\ \mbox{\tiny{C}}}^{\mbox{\tiny{A}}}=\mathrm {\imath}\left(
\im\sqrt{\sigma}\delta_{\phantom\
\mbox{\tiny{C}}}^{\mbox{\tiny{A}}}-\frac{\epsilon}{2
\im\sqrt{\sigma}} s^{\mbox{\tiny{A}}}s_{\mbox{\tiny{C}}}\right)\ .
\end{equation}
For   $\|\mathbf{s}\|^{2} \neq0 $ we obtain
\begin{align}
\label{per-ca}
\alpha&=\imath\,\frac{-\im\sqrt{\sigma}\mp\sqrt{(
\im\sqrt{\sigma})^{2}-\epsilon\|\mathbf{s}\|^{2} }}{\|\mathbf{s}\|^{2}}
\\
\nonumber \textrm{and}&\quad\re\sqrt{\sigma}=0\quad\textrm{for}\quad
\sigma+\epsilon\|\mathbf{s}\|^{2}<0 \quad\textrm{where}\quad
\sigma=- (\im\sqrt{\sigma})^{2}.
%&(\im\sqrt{\sigma})^{2}=-\sigma\\
\end{align}
% (69) in (70), the final expression
Finally, using  (\ref{per-ca}) in  (\ref{plo-ne}) we
find the following expression for \ $\phi_{\phantom\
\mbox{\tiny{C}}}^{\mbox{\tiny{A}}}$:
\begin{eqnarray}
\nonumber\phi_{\phantom\ \mbox{\tiny{C}}}^{\mbox{\tiny{A}}}&=&
\mathrm {\imath}\left[ \im\sqrt{\sigma}\delta_{\phantom\
\mbox{\tiny{C}}}^{\mbox{\tiny{A}}}-\|\mathbf{s}\|^{-2}\left(\im\sqrt{\sigma}\pm\sqrt{(
\im\sqrt{\sigma})^{2}-\epsilon\|\mathbf{s}\|^{2} }\right)
s^{\mbox{\tiny{A}}}s_{\mbox{\tiny{C}}}\right]\\
\label{sc-la}&&\mbox{for}\quad
\sigma+\epsilon\|\mathbf{s}\|^{2} <0
\end{eqnarray}
Thus, in particular we find
\begin{eqnarray}
\nonumber\phi_{\phantom\ \mbox{\tiny{C}}}^{\mbox{\tiny{A}}}&=&
\mathrm {\imath}\left[ \im\sqrt{\sigma}\delta_{\phantom\
\mbox{\tiny{C}}}^{\mbox{\tiny{A}}}-\|\mathbf{s}\|^{-2}\left(\im\sqrt{\sigma}\pm\sqrt{(
\im\sqrt{\sigma})^{2}+\|\mathbf{s}\|^{2} }\right)
s^{\mbox{\tiny{A}}}s_{\mbox{\tiny{C}}}\right]\\
\label{tir-te1}&&\mbox{for}\quad \epsilon=-1 \quad\mbox{and}\quad
\|\mathbf{s}\|^{2} -\sigma>0
\end{eqnarray}
and
\begin{eqnarray}
\nonumber\phi_{\phantom\ \mbox{\tiny{C}}}^{\mbox{\tiny{A}}}&=&
\mathrm {\imath}\left[ \im\sqrt{\sigma}\delta_{\phantom\
\mbox{\tiny{C}}}^{\mbox{\tiny{A}}}-\|\mathbf{s}\|^{-2}\left(\im\sqrt{\sigma}\pm\sqrt{(
\im\sqrt{\sigma})^{2}-\|\mathbf{s}\|^{2} }\right)
s^{\mbox{\tiny{A}}}s_{\mbox{\tiny{C}}}\right]\\
\label{tir-te2}&&\mbox{for}\quad \epsilon=+1 \quad\mbox{and}\quad
\|\mathbf{s}\|^{2} <-\sigma
\end{eqnarray}

 While for
\begin{equation}\label{Cabibbo-KM}
\sigma+\epsilon\|\mathbf{s}\|^{2}>0
\end{equation}
we find:
\begin{align*}
\alpha&=\mp\frac{\sqrt{\epsilon\|\mathbf{s}\|^{2}-(
\im\sqrt{\sigma})^{2}}}{\|\mathbf{s}\|^{2}}-\imath\,\frac{\im\sqrt{\sigma}}{\|\mathbf{s}\|^{2}}
\quad\textrm{and}\quad\re\sqrt{\sigma}=0.
\end{align*}
Finally  the matrix $\phi_{\phantom\
\mbox{\tiny{C}}}^{\mbox{\tiny{A}}}$ in this case is
\begin{align}\label{noduli1}
\phi_{\phantom\
\mbox{\tiny{C}}}^{\mbox{\tiny{A}}}&=\mp\frac{\sqrt{\epsilon\|\mathbf{s}\|^{2}-(
\im\sqrt{\sigma})^{2}}}{\|\mathbf{s}\|^{2}} \
s^{\mbox{\tiny{A}}}s_{\mbox{\tiny{C}}}+ \mathrm {\imath}\left(
\im\sqrt{\sigma}\delta_{\phantom\
\mbox{\tiny{C}}}^{\mbox{\tiny{A}}}-\frac{\im\sqrt{\sigma}}{\|\mathbf{s}\|^{2}}
s^{\mbox{\tiny{A}}}s_{\mbox{\tiny{C}}}\right).
\end{align}
Thus, in particular we have:
\begin{align}\nonumber
\phi_{\phantom\
\mbox{\tiny{C}}}^{\mbox{\tiny{A}}}&=\mp\frac{\sqrt{-\left[\|\mathbf{s}\|^{2}+(
\im\sqrt{\sigma})^{2}\right]}}{\|\mathbf{s}\|^{2}} \
s^{\mbox{\tiny{A}}}s_{\mbox{\tiny{C}}}+ \mathrm {\imath}\left(
\im\sqrt{\sigma}\delta_{\phantom\
\mbox{\tiny{C}}}^{\mbox{\tiny{A}}}-\frac{\im\sqrt{\sigma}}{\|\mathbf{s}\|^{2}}
s^{\mbox{\tiny{A}}}s_{\mbox{\tiny{C}}}\right)\\
\label{term1} \mbox{for}& \quad\epsilon=-1
\quad\mbox{and}\quad\|\mathbf{s}\|^{2}<\sigma
\end{align}
and
\begin{align}\nonumber
\phi_{\phantom\
\mbox{\tiny{C}}}^{\mbox{\tiny{A}}}&=\mp\frac{\sqrt{\|\mathbf{s}\|^{2}-(
\im\sqrt{\sigma})^{2}}}{\|\mathbf{s}\|^{2}} \
s^{\mbox{\tiny{A}}}s_{\mbox{\tiny{C}}}+ \mathrm {\imath}\left(
\im\sqrt{\sigma}\delta_{\phantom\
\mbox{\tiny{C}}}^{\mbox{\tiny{A}}}-\frac{\im\sqrt{\sigma}}{\|\mathbf{s}\|^{2}}
s^{\mbox{\tiny{A}}}s_{\mbox{\tiny{C}}}\right)\\
\label{ter-2} \mbox{for}& \quad\epsilon=+1
\quad\mbox{and}\quad\|\mathbf{s}\|^{2}>-\sigma
\end{align}
%
%\end{description}
%\end{description}
%
 We summarize the  complex solutions of Eq.\il
(\ref{spa-rti}) for $\sigma<0$
in Table\il\ref{al-be}  and for  $\sigma>0$  in Table\il\ref{tab:occ}.
\begin{table}[ht]
%\begin{center}
\resizebox{1.5\textwidth}{!}{%
\begin{tabular}{ccc}
\toprule
 $\sigma<0$&$\epsilon=-1$ &$\epsilon=+1 $
\\   \midrule
\bfseries$\|\mathbf{s}\|^{2}<-|\sigma|$ &$\phi_{\phantom\
\mbox{\tiny{C}}}^{\mbox{\tiny{A}}}=\mp\|\mathbf{s}\|^{-2}\sqrt{-\|\mathbf{s}\|^{2}-(
\mbox{Im}\sqrt{\sigma})^{2}} s^{\mbox{\tiny{A}}}s_{\mbox{\tiny{C}}}+
\mathrm {\imath}\left(\mbox{Im}\sqrt{\sigma}\delta_{\phantom\
\mbox{\tiny{C}}}^{\mbox{\tiny{A}}}-\frac{\mbox{Im}\sqrt{\sigma}}{\|\mathbf{s}\|^{2}}
s^{\mbox{\tiny{A}}}s_{\mbox{\tiny{C}}}\right)$&$\phi_{\phantom\
\mbox{\tiny{C}}}^{\mbox{\tiny{A}}}= \mathrm {\imath}\left[
\mbox{Im}\sqrt{\sigma}\delta_{\phantom\
\mbox{\tiny{C}}}^{\mbox{\tiny{A}}}-\|\mathbf{s}\|^{-2}\left(\mbox{Im}\sqrt{\sigma}\pm\sqrt{(
\mbox{Im}\sqrt{\sigma})^{2}-\|\mathbf{s}\|^{2} }\right)
s^{\mbox{\tiny{A}}}s_{\mbox{\tiny{C}}}\right]$
\\
\bfseries$\|\mathbf{s}\|^{2}=-|\sigma|$ &$\phi_{\phantom\
\mbox{\tiny{C}}}^{\mbox{\tiny{A}}}= \mathrm
{\imath}\left(\mbox{Im}\sqrt{\sigma}\delta_{\phantom\
\mbox{\tiny{C}}}^{\mbox{\tiny{A}}}+\frac{1}{\mbox{Im}\sqrt{\sigma}}
s^{\mbox{\tiny{A}}}s_{\mbox{\tiny{C}}}\right)$&$\phi_{\phantom\
\mbox{\tiny{C}}}^{\mbox{\tiny{A}}}= \mathrm {\imath}\left[
\mbox{Im}\sqrt{\sigma}\delta_{\phantom\
\mbox{\tiny{C}}}^{\mbox{\tiny{A}}}+\frac{1}{\mbox{Im}\sqrt{\sigma}}(1\pm\sqrt{2})
s^{\mbox{\tiny{A}}}s_{\mbox{\tiny{C}}}\right]$\\
\\
\bfseries$\|\mathbf{s}\|^{2}\in]-|\sigma|,0[$ &$\phi_{\phantom\
\mbox{\tiny{C}}}^{\mbox{\tiny{A}}}= \mathrm {\imath}\left[
\mbox{Im}\sqrt{\sigma}\delta_{\phantom\
\mbox{\tiny{C}}}^{\mbox{\tiny{A}}}-\|\mathbf{s}\|^{-2}\left(\mbox{Im}\sqrt{\sigma}\pm\sqrt{(
\mbox{Im}\sqrt{\sigma})^{2}+\|\mathbf{s}\|^{2} }\right)
s^{\mbox{\tiny{A}}}s_{\mbox{\tiny{C}}}\right]$&$\phi_{\phantom\
\mbox{\tiny{C}}}^{\mbox{\tiny{A}}}= \mathrm {\imath}\left[
\mbox{Im}\sqrt{\sigma}\delta_{\phantom\
\mbox{\tiny{C}}}^{\mbox{\tiny{A}}}-\|\mathbf{s}\|^{-2}\left(\mbox{Im}\sqrt{\sigma}\pm\sqrt{(
\mbox{Im}\sqrt{\sigma})^{2}-\|\mathbf{s}\|^{2} }\right)
s^{\mbox{\tiny{A}}}s_{\mbox{\tiny{C}}}\right]$\\
\\
\\
\bfseries$\|\mathbf{s}\|^{2}=0$ &$\phi_{\phantom\
\mbox{\tiny{C}}}^{\mbox{\tiny{A}}}=\mathrm {\imath}\left(
\mbox{Im}\sqrt{\sigma}\delta_{\phantom\
\mbox{\tiny{C}}}^{\mbox{\tiny{A}}}+\frac{1}{2
\mbox{Im}\sqrt{\sigma}}
s^{\mbox{\tiny{A}}}s_{\mbox{\tiny{C}}}\right)
 $ &$\phi_{\phantom\
\mbox{\tiny{C}}}^{\mbox{\tiny{A}}}=\mathrm {\imath}\left(
\mbox{Im}\sqrt{\sigma}\delta_{\phantom\
\mbox{\tiny{C}}}^{\mbox{\tiny{A}}}-\frac{1}{2
\mbox{Im}\sqrt{\sigma}}
s^{\mbox{\tiny{A}}}s_{\mbox{\tiny{C}}}\right)
$\\
\\
\bfseries$\|\mathbf{s}\|^{2}	\in]0,\sigma|[$ &$\phi_{\phantom\
\mbox{\tiny{C}}}^{\mbox{\tiny{A}}}= \mathrm {\imath}\left[
\mbox{Im}\sqrt{\sigma}\delta_{\phantom\
\mbox{\tiny{C}}}^{\mbox{\tiny{A}}}-\|\mathbf{s}\|^{-2}\left(\mbox{Im}\sqrt{\sigma}\pm\sqrt{(
\mbox{Im}\sqrt{\sigma})^{2}+\|\mathbf{s}\|^{2} }\right)
s^{\mbox{\tiny{A}}}s_{\mbox{\tiny{C}}}\right] $&$\phi_{\phantom\
\mbox{\tiny{C}}}^{\mbox{\tiny{A}}}= \mathrm {\imath}\left[
\mbox{Im}\sqrt{\sigma}\delta_{\phantom\
\mbox{\tiny{C}}}^{\mbox{\tiny{A}}}-\|\mathbf{s}\|^{-2}\left(\mbox{Im}\sqrt{\sigma}\pm\sqrt{(
\mbox{Im}\sqrt{\sigma})^{2}-\|\mathbf{s}\|^{2} }\right)
s^{\mbox{\tiny{A}}}s_{\mbox{\tiny{C}}}\right]
$\\
\\
$\|\mathbf{s}\|^{2}=|\sigma|$ &$\phi_{\phantom\
\mbox{\tiny{C}}}^{\mbox{\tiny{A}}}= \mathrm {\imath}\left[
\mbox{Im}\sqrt{\sigma}\delta_{\phantom\
\mbox{\tiny{C}}}^{\mbox{\tiny{A}}}-\frac{1}{\mbox{Im}\sqrt{\sigma}}(1\pm\sqrt{2})
s^{\mbox{\tiny{A}}}s_{\mbox{\tiny{C}}}\right] $&$\phi_{\phantom\
\mbox{\tiny{C}}}^{\mbox{\tiny{A}}}= \mathrm {\imath}\left[
\mbox{Im}\sqrt{\sigma}\delta_{\phantom\
\mbox{\tiny{C}}}^{\mbox{\tiny{A}}}-\frac{1}{\mbox{Im}\sqrt{\sigma}}
s^{\mbox{\tiny{A}}}s_{\mbox{\tiny{C}}}\right]
$\\
\\
\bfseries$\|\mathbf{s}\|^{2}>|\sigma|$&$\phi_{\phantom\
\mbox{\tiny{C}}}^{\mbox{\tiny{A}}}= \mathrm {\imath}\left[
\mbox{Im}\sqrt{\sigma}\delta_{\phantom\
\mbox{\tiny{C}}}^{\mbox{\tiny{A}}}-\|\mathbf{s}\|^{-2}\left(\mbox{Im}\sqrt{\sigma}\pm\sqrt{(
\mbox{Im}\sqrt{\sigma})^{2}+\|\mathbf{s}\|^{2} }\right)
s^{\mbox{\tiny{A}}}s_{\mbox{\tiny{C}}}\right]
$&$\phi_{\mbox{\tiny{C}}}^{\mbox{\tiny{A}}}=\mp\|\mathbf{s}\|^{-2}\sqrt{\|\mathbf{s}\|^{2}-(
\mbox{Im}\sqrt{\sigma})^{2}} \
s^{\mbox{\tiny{A}}}s_{\mbox{\tiny{C}}}+ \mathrm
{\imath}\left(\mbox{Im}\sqrt{\sigma}\delta_{\phantom\
\mbox{\tiny{C}}}^{\mbox{\tiny{A}}}-\frac{\mbox{Im}\sqrt{\sigma}}{\|\mathbf{s}\|^{2}}
s^{\mbox{\tiny{A}}}s_{\mbox{\tiny{C}}}\right) $
\\
\\\bottomrule
\end{tabular}
}
%\end{center}
\centering \caption{Deforming matrices for three--dimensional metrics.
$2^{\circ}$ case: Complex solutions with $\sigma<0$. In this case we
find $\mbox{Re}\sqrt{\sigma}=0$ and $|\sigma|=\left( \mbox{Im}
\sqrt{\sigma}\right)^{2}$.}\label{al-be}
\end{table}
%First, let us note that, for $\sigma>0$ and
%%\toprule, \midrule e \bottomrule
%
\begin{table}[h]
%\begin{center}
\resizebox{1.17\textwidth}{!}{%
\begin{tabular}{ccc}
\toprule $\sigma>0$&$\epsilon=-1$ &$\epsilon=+1 $
\\   \midrule
\bfseries$\|\mathbf{s}\|^{2}<-\sigma$ &$\ \checkmark
$&$\phi_{\phantom\
\mbox{\tiny{C}}}^{\mbox{\tiny{A}}}=\sqrt{\sigma}\delta_{\phantom\
\mbox{\tiny{C}}}^{\mbox{\tiny{A}}}-\frac{\sqrt{\sigma}}{\|\mathbf{s}\|^{2}}
s^{\mbox{\tiny{A}}}s_{\mbox{\tiny{C}}}\mp\frac{\mathrm
{\imath}}{\|\mathbf{s}\|^{2}}\sqrt{-(\sigma+\|\mathbf{s}\|^{2})}
s^{\mbox{\tiny{A}}}s_{\mbox{\tiny{C}}} $\\
\\
\bfseries$\|\mathbf{s}\|^{2}=-\sigma$ &$\checkmark$&$\phi_{\phantom\
\mbox{\tiny{C}}}^{\mbox{\tiny{A}}}=\sqrt{\sigma}\delta_{\phantom\
\mbox{\tiny{C}}}^{\mbox{\tiny{A}}}+\frac{1}{\sqrt{\sigma}}
s^{\mbox{\tiny{A}}}s_{\mbox{\tiny{C}}} $\\
\\
\bfseries$\|\mathbf{s}\|^{2}=0$ &$\phi_{\phantom\
\mbox{\tiny{C}}}^{\mbox{\tiny{A}}}=\sqrt{\sigma}-\frac{1}{2\sqrt{\sigma}}s^{\mbox{\tiny{A}}}s_{\mbox{\tiny{C}}}
$ &$\phi_{\phantom\
\mbox{\tiny{C}}}^{\mbox{\tiny{A}}}=\sqrt{\sigma}+\frac{1}{2\sqrt{\sigma}}s^{\mbox{\tiny{A}}}s_{\mbox{\tiny{C}}} $\\
\\
$\|\mathbf{s}\|^{2}=\sigma$ &$\phi_{\phantom\
\mbox{\tiny{C}}}^{\mbox{\tiny{A}}}=\sqrt{\sigma}\delta_{\phantom\
\mbox{\tiny{C}}}^{\mbox{\tiny{A}}}-\frac{1}{\sqrt{\sigma}}
s^{\mbox{\tiny{A}}}s_{\mbox{\tiny{C}}}$ &$\checkmark$\\
\\
\bfseries$\|\mathbf{s}\|^{2}>\sigma$ &$\phi_{\phantom\
\mbox{\tiny{C}}}^{\mbox{\tiny{A}}}=\sqrt{\sigma}\delta_{\phantom\
\mbox{\tiny{C}}}^{\mbox{\tiny{A}}}-\frac{\sqrt{\sigma}}{\|\mathbf{s}\|^{2}}
s^{\mbox{\tiny{A}}}s_{\mbox{\tiny{C}}}\mp\frac{\mathrm
{\imath}}{\|\mathbf{s}\|^{2}}\sqrt{-(\sigma-\|\mathbf{s}\|^{2})}
s^{\mbox{\tiny{A}}}s_{\mbox{\tiny{C}}}$&$\checkmark $
\\
\\\bottomrule
\end{tabular}}
%\end{center}
\caption{Deforming matrices for three--dimensional metrics. $2^{\circ}$
case: Complex solutions with $\sigma>0$.}\label{tab:occ}
\end{table}
%\mbox{Im}
%
%\newcolumntype{U}{>{\centering\arraybackslash$}X<{$}}
%\begin{tabularx}{0.9\columnwidth}{*{7}{U}}
%
\section{The causal structure}\label{Sec:thecausalstructure}
In this section we investigate  the casual structure of a deformed
manifold compared with the original one. In order to start our
considerations, we show how, for $h$, $\sigma$ and $\mathbf{s}$
fixed, the norm of tangent vectors on the manifold  $M$ changes for deformation of
the metric tensors.
Thus, we consider the norm $\| \mathbf{A}\|^{2}_{g}$ of
a vector  $ \mathbf{A}$, defined by the deformed metric $g$ and the
$\| \mathbf{A}\|^{2}_{ h}$ defined from the metric $h$ as follows:
\begin{eqnarray}
\|\mathbf{A}\|^{2}_{g}&\equiv&g_ {\mu\nu } A^{\mu}A^{\nu}, \qquad
\|\mathbf{A}\|^{2}_{ h}\equiv h_{\mu\nu}A^{\mu}A^{\nu}.
\end{eqnarray}
%and also their expressions with index notation
In the following, for simplicity we will use the notation
\begin{equation}\label{not-z}
(sA)^{2} \equiv s_{\mu}A^{\mu}s_{\nu}A^{\nu}.
\end{equation}
Therefore, writing the norms as:
\begin{equation}\label{three--dim}
g_ {\mu\nu } A^{\mu}A^{\nu}= (\sigma h_{\mu\nu}+ \epsilon
s_{\mu}s_{\nu})A^{\mu}A^{\nu}
\end{equation}
we have
\begin{eqnarray}
\label{K-s-me} (\sigma h_{\mu\nu}+ \epsilon
s_{\mu}s_{\nu})A^{\mu}A^{\nu}&=&\sigma
h_{\mu\nu}A^{\mu}A^{\nu}+\epsilon
s_{\mu}A^{\mu}s_{\nu}A^{\nu},\\\label{riu-gr}
\|\mathbf{A}\|^{2}_{g}&=& \sigma\|\mathbf{A}\|^{2}_{
h}+\epsilon(\mathbf{s}A)^{2}.
\end{eqnarray}
Moreover, at (\ref{riu-gr})  a scalar
relation $\Psi(\sigma,\|\mathbf{s}\|)=0$ between $\sigma$ and
$\mathbf{s}$ should be imposed.
Below we describe the various cases we find from  this decomposition. We can distinguish the cases in which the
casual structure remains unchanged and the cases in which it is locally
and globally  changed.
\begin{description}
\item[1$^{\circ}$ caso:] $\mathbf{s}=0$.
In this first case    transformation (\ref{eqn:eulero}) is  a conformal
deformation, thus
$$
\|\mathbf{A}\|_{g}^{2}=\sigma\|\mathbf{A}\|^{2}_{ h}.
$$
%With this definitions in mind, let us consider
We note that if $\sigma$ is a generic  function,   where this
is positive the casual structure is preserved, while where it is
negative the casual structure is changed, and  if the function
is continue, it will be zero  in some regions, thus there will be a
singularity. This pathologic situation could be avoided if we
consider only strictly positive scalar fields $\sigma$. In this case
the casual structure will be preserved in the entire  spacetime.
%The inverse of $ tensore metric \ g (tilde) _ () $ ij is
%%
%$ $ RJ ^ (h) \ Equiv \ phi ^ (-1) \ left (g ^ rj) (+ \ frac (1) (\

%Consider, then, the norm squared $ \ | \ mathbf (A) \ | ^ (2) _ (g)
%  Will be \ `(a) preserved in every point of space.
\item [2$^{\circ}$ case:]    $ \mathbf{s}\neq \mathbf{0} $
Let us split this second case into two subcases.

Let be $\mathbf{A}\neq \mathbf{0}$ a vector, if $ s_{\mu}A^{\mu}=0 $
then $ \|\mathbf{A}\|_{g}^{2}\equiv\sigma\|\mathbf{A}\|^{2}_{ h}$,
or we have  $s_{\mu}A^{\mu}\neq0$.

If $\mathbf{A}$ and $\mathbf{s}$ are real vectors we have
\begin{equation}\label{strettposit}
(\mathbf{s}\mathbf{A})^{2} > 0.
\end{equation}
From the relation
\begin{equation}
\|\mathbf{A}\|_{g}^{2}=\sigma\|\mathbf{A}\|^{2}_{
h}+\epsilon(\mathbf{s}\mathbf{A})^{2}
\end{equation}
we have the following possibilities
\begin{enumerate}
\item If $\sigma\|\mathbf{A}\|^{2}_{ h}>(\mathbf{s}\mathbf{A})^{2}$ then
\begin{equation}\label{Mi-ak-od}
\|\mathbf{A}\|_{g}^{2} > 0\quad \mbox{for}\quad  \epsilon=\pm1,
\end{equation}
\\
%%
%\begin{eqnarray}\label{arrivo}
%\textrm{for}\quad\epsilon=+1:&&\|\mathbf{A}\|_{g}^{2} > 0\\
%\textrm{for}\quad\epsilon=-1:&&\|\mathbf{A}\|_{g}^{2}>0
%\end{eqnarray}
%can be written as follows
\item If  $\sigma\|\mathbf{A}\|^{2}_{
h}=(\mathbf{s}\mathbf{A})^{2}$we have:
\begin{subequations}
\begin{eqnarray}
\|\mathbf{A}\|_{g}^{2}>0&\quad&\textrm{for}\quad \epsilon=+1 ,\\
\|\mathbf{A}\|_{g}^{2}=0&\quad&\textrm{for}\quad\epsilon=-1 ,
\end{eqnarray}
\end{subequations}
\item  If $\sigma\|\mathbf{A}\|^{2}_{
h}<(\mathbf{s}\mathbf{A})^{2}$ then we have the following three
possibilities:

$ \mbox{If}\quad 0<\sigma\|\mathbf{A}\|^{2}_{
h}<(\mathbf{s}\mathbf{A})^{2} $
\begin{subequations}
\begin{eqnarray}
\|\mathbf{A}\|_{g}^{2}>0\quad&\textrm{for}&\epsilon=+1\ , \ \\
\|\mathbf{A}\|_{g}^{2}<0\quad&\textrm{for}&\epsilon=-1\ .
\end{eqnarray}
\end{subequations}
If $\sigma\|\mathbf{A}\|^{2}_{ h}=0$ instead:
\begin{subequations}
\begin{eqnarray}
\|\mathbf{A}\|_{g}^{2}>0 \quad&\textrm{for}&\epsilon=+1\ , \\
\|\mathbf{A}\|_{g}^{2}<0\quad&\textrm{for}&\epsilon=-1\ .
\end{eqnarray}
\end{subequations}
%Table I lists
%considered are listed in Table I. The exact forms of a(t) at all the
%critical points are summarised in the table 4.1.
%5.1 summarises the results found in this section.
If $\|\mathbf{A}\|^{2}_{\sigma h}<0$   we finally find the results
listed in  Table\il\ref{tab:grosse}, where $|
\sigma\|\mathbf{A}\|^{2}_{ h}|$ is the module of
$\sigma\|\mathbf{A}\|^{2}_{ h}$.
%
%{||p{5 cm }||*{2}{c|}|}
\begin{table}[hbpt]
\begin{center}
\resizebox{.7\textwidth}{!}{%
\begin{tabular}{ccc}
\toprule $ \sigma \|\mathbf{A}\|^{2}_{h}<0$ &$\epsilon=-1$
&$\epsilon=+1 $
\\  \midrule
$|\sigma\|\mathbf{A}\|^{2}_{ h}|>(\mathbf{s}\mathbf{A})^{2}$
&$\|\mathbf{A}\|_{g}^{2}<0$ &$\|\mathbf{A}\|_{g}^{2}<0$
\\
\bfseries$|\sigma\|\mathbf{A}\|^{2}_{
h}|<(\mathbf{s}\mathbf{A})^{2}$ &$\|\mathbf{A}\|_{g}^{2}<0$
&$\|\mathbf{A}\|_{g}^{2}>0$
\\
\bfseries$|\sigma\|\mathbf{A}\|^{2}_{
h}|=(\mathbf{s}\mathbf{A})^{2}$
&$\|\mathbf{A}\|_{g}^{2}<0$&$\|\mathbf{A}\|_{g}^{2}=0$
\\
\bottomrule
\end{tabular}
}
\end{center}
\caption{Casual structure of a deformed manifold: case
$\|\mathbf{A}\|^{2}_{\sigma h}<0$.}\label{tab:grosse}
\end{table}
In the Table\il\ref{tab:Q-J} we report the results obtained
for $(\mathbf{s}\mathbf{A})^{2}>0$.
\begin{table}[hbpt]
\begin{center}
\begin{tabular}{ccc}
\toprule
 $(\mathbf{s}\mathbf{A})^{2}>0
$&$\epsilon=-1$&$\epsilon=+1$
\\
\midrule \bfseries$\sigma\|\mathbf{A}\|^{2}_{
h}>(\mathbf{s}\mathbf{A})^{2}$
&$\|\mathbf{A}\|_{g}^{2}>0$&$\|\mathbf{A}\|_{g}^{2}>0$
\\
\\
\bfseries$\sigma\|\mathbf{A}\|^{2}_{ h}=(\mathbf{s}\mathbf{A})^{2}$
&$\|\mathbf{A}\|_{g}^{2}=0$&$\|\mathbf{A}\|_{g}^{2}=2\sigma\|\mathbf{A}\|^{2}_{
h}$
\\
%\hlinetable
\\
\bfseries$\sigma\|\mathbf{A}\|^{2}_{h}\in]-(\mathbf{s}\mathbf{A})^{2},(\mathbf{s}\mathbf{A})^{2}[$
&$\|\mathbf{A}\|_{g}^{2}<0$ &$\|\mathbf{A}\|_{g}^{2}>0$
\\
%\hline
\\
\bfseries$\sigma\|\mathbf{A}\|^{2}_{ h}=0$
&$\|\mathbf{A}\|_{g}^{2}=-(\mathbf{s}\mathbf{A})^{2}$&$\|\mathbf{A}\|_{g}^{2}=(\mathbf{s}\mathbf{A})^{2}$
\\
%\hline
\\
\bfseries$\sigma\|\mathbf{A}\|^{2}_{ h}=-(\mathbf{s}\mathbf{A})^{2}$
&$\|\mathbf{A}\|_{g}^{2}=-2\sigma\|\mathbf{A}\|^{2}_{
h}$&$\|\mathbf{A}\|_{g}^{2}=0$
\\
%\hline
\\
$\sigma\|\mathbf{A}\|^{2}_{ h}<-(\mathbf{s}\mathbf{A})^{2}$
&$\|\mathbf{A}\|_{g}^{2}<0$ &$\|\mathbf{A}\|_{g}^{2}<0$
\\
\hline
\bottomrule
\end{tabular}
\end{center}
\caption{Casual structure of a deformed manifold. Case
 $(\mathbf{s}\mathbf{A})^{2}>0 $}\label{tab:Q-J}
\end{table}
\end{enumerate}
\end{description}
We note that   the case
$(\mathbf{s}\mathbf{A})^{2}<0$ should be avoided because
Eq.\il(\ref{eqn:eulero}) is  proved  only in the analytical case,
where all the data are real analytic functions, however
since the case of complex $\mathbf{s}$ vector  occurs from
deformation (\ref{Mas-tt}) parameterized by scalar fields, this case
 is also  studied. We summarize the results for the
$(\mathbf{s}\mathbf{A})^{2}<0$  case in  Table\il\ref{tab:Q-J1}.
\begin{table}[hbpt]
\begin{center}
\begin{tabular}{ccc}
\toprule $(\mathbf{s}\mathbf{A})^{2}<0 $ &$\epsilon=-1$
&$\epsilon=+1$
\\
\midrule \bfseries$\sigma\|\mathbf{A}\|^{2}_{
h}<(\mathbf{s}\mathbf{A})^{2}$ &$\|\mathbf{A}\|_{g}^{2}<0$
&$\|\mathbf{A}\|_{g}^{2}<0$
\\
\\
\bfseries$\sigma\|\mathbf{A}\|^{2}_{ h}=(\mathbf{s}\mathbf{A})^{2}$
&$\|\mathbf{A}\|_{g}^{2}=0$&$\|\mathbf{A}\|_{g}^{2}=2\sigma\|\mathbf{A}\|^{2}_{
h}$
\\
%\hline
\\
\bfseries$\sigma\|\mathbf{A}\|^{2}_{h}\in](\mathbf{s}\mathbf{A})^{2},-(\mathbf{s}\mathbf{A})^{2}[$
&$\|\mathbf{A}\|_{g}^{2}>0$ &$\|\mathbf{A}\|_{g}^{2}<0$
\\
%\hline
\\
\bfseries$\sigma\|\mathbf{A}\|^{2}_{ h}=0$
&$\|\mathbf{A}\|_{g}^{2}=-(\mathbf{s}\mathbf{A})^{2}$&$\|\mathbf{A}\|_{g}^{2}=(\mathbf{s}\mathbf{A})^{2}$
\\
%\hline
\\
\bfseries$\sigma\|\mathbf{A}\|^{2}_{ h}=-(\mathbf{s}\mathbf{A})^{2}$
&$\|\mathbf{A}\|_{g}^{2}=-2\sigma\|\mathbf{A}\|^{2}_{
h}$&$\|\mathbf{A}\|_{g}^{2}=0$
\\
%\hline
\\
$\sigma\|\mathbf{A}\|^{2}_{ h}>-(\mathbf{s}\mathbf{A})^{2}$
&$\|\mathbf{A}\|_{g}^{2}>0$ &$\|\mathbf{A}\|_{g}^{2}>0$
\\
\bottomrule
\end{tabular}
\end{center}
\caption{Casual structure of a deformed manifold. Case
 $(\mathbf{s}\mathbf{A})^{2}<0 $}\label{tab:Q-J1}
\end{table}
\section{Some deformed  three--dimensional metrics}\label{SeC:example}
In this section we  investigate the  local deformations  of some
three--dimensional Riemannian manifolds  into flat metrics. We start
by applying the result (\ref{eq}) to  the particular simple cases of
constant curvature metrics studied in \cite{three--dim}: the metric of a
three--dimensional sphere, and the metric of a Kerr space. Then, by using
the results listed in Table\il\ref{tab:minute} and
\ref{al-be}, we construct the  deforming matrices associated to
these deformed metrics.
\subsection{The three--dimensional sphere}\label{Sec:3-sferical}
We  consider the three--dimensional sphere   $S^{3}$. The metric for
the  three--manifold  in Schwarzschild spacetime,
 is\footnote{In units of $c=G=1.$}:
\begin{equation}\label{Per-dri}
\hat{g}=\left(1-\frac{2m}{r}\right)^{-1}dr\otimes
dr+r^{2}\left(d\theta\otimes d\theta+\sin\theta^{2}d\phi\otimes
d\phi\right)
\end{equation}
 in the polar coordinate
$\left(r,\theta,\phi\right)$
where $m$ is the body mass and the metric is considered in the
region $r
> 2m$, \cite{wald,Kramer-ste-M}.

Metric (\ref{Per-dri}) can be locally deformed into a flat metric in several
ways. For example in \cite{three--dim}  the metric
$\hat{g}$ is obtained by the deformation of a metric $\tilde{g}$ by
the (\ref{eqn:eulero}) as:
\begin{equation}\label{eqn:eulero26}
\hat{g}=\tilde{g}+ \textbf{s}\otimes\textbf{s}\ ,
\end{equation}
where $\sigma=1$, $\epsilon=+1$ and  $\mathbf{s}=\sqrt{k^{-1}-1} dr$
where $k\equiv\left(1-\frac{2m}{r}\right)$ and
\begin{equation}\label{Per-dri1}
\tilde{g}=dr\otimes dr+r^{2}\left(d\theta\otimes
d\theta+\sin\theta^{2}d\phi\otimes d\phi\right).
\end{equation}
We can write $\tilde{g}$ as
\begin{equation}\label{Per-dri2}
\tilde{g}=\omega^{\hat{r}}\otimes
\omega^{\hat{r}}+\omega^{\hat{\theta}}\otimes
\omega^{\hat{\theta}}+\omega^{\hat{\phi}}\otimes
\omega^{\hat{\phi}},
\end{equation}
in the triad $\omega^{\mbox{\tiny{A}}}_{a}$
\begin{equation}\label{Val-chi-ra}
\omega^{\hat{r}}=dr,\quad \omega^{\hat{\theta}}=r
d\theta,\quad\omega^{\hat{\phi}}=r \sin\theta d\phi
\end{equation}
%%as
while  the metric $\hat{g}$ reads
\begin{equation}\label{PD3}
\hat{g}=\frac{1}{k}\omega^{\hat{r}}\otimes
\omega^{\hat{r}}+\omega^{\hat{\theta}}\otimes
\omega^{\hat{\theta}}+\omega^{\hat{\phi}}\otimes \omega^{\hat{\phi}}
\end{equation}
The  deforming matrix $\phi_{\phantom\
\mbox{\tiny{C}}}^{\mbox{\tiny{A}}}$ associated to the deformation
(\ref{eqn:eulero26}) has the form (\ref{per-ta}), or
\begin{equation}\label{perinbo}
\phi_{\phantom\
\mbox{\tiny{C}}}^{\mbox{\tiny{A}}}=\sqrt{\sigma}\delta_{\phantom\
\mbox{\tiny{C}}}^{\mbox{\tiny{A}}}+\alpha
s^{\mbox{\tiny{A}}}s_{\mbox{\tiny{C}}} \, ,
\end{equation}
where in this case  $\sigma=1$ and
$\|\mathbf{s}\|^{2}_{\tilde{g}}\equiv\left(1/k-1\right)>0$.

In particular   equation (\ref{eq}) has the following solutions:
\begin{eqnarray}
\label{pan-for-1}\phi_{\phantom\
\mbox{\tiny{C}}}^{\mbox{\tiny{A}}}&=&\sqrt{\sigma}\delta_{\phantom\
\mbox{\tiny{C}}}^{\mbox{\tiny{A}}}+
\frac{-\sqrt{\sigma}\mp\sqrt{\sigma+\|\mathbf{s}\|^{2}}}{\|\mathbf{s}\|^{2}}
\mathbf{s}^{\mbox{\tiny{A}}}\mathbf{s}_{\mbox{\tiny{C}}},\quad
\mbox{for}\quad \|\mathbf{s}\|^{2}_{\tilde{g}}\in\left]0,1\right[
\end{eqnarray}
and substituting in the (\ref{pan-for-1})
$\|\mathbf{s}\|^{2}_{\tilde{g}}$ and $\sigma$ we have:
\begin{eqnarray}
\label{Pianoforte2}\phi_{\phantom\
\mbox{\tiny{C}}}^{\mbox{\tiny{A}}}&=&\delta_{\phantom\
\mbox{\tiny{C}}}^{\mbox{\tiny{A}}}-
\left(1\pm\frac{1}{\sqrt{k}}\right)\frac{1}{\left(\frac{1}{k}-1\right)}
\mathbf{s}^{\mbox{\tiny{A}}}\mathbf{s}_{\mbox{\tiny{C}}},\quad
\mbox{for}\quad k\in\left]1/2,1\right[
\end{eqnarray}
or also
\begin{eqnarray}
\label{Pianoforte3}\phi_{\phantom\
\mbox{\tiny{C}}}^{\mbox{\tiny{A}}}&=&\delta_{\phantom\
\mbox{\tiny{C}}}^{\mbox{\tiny{A}}}-
\frac{r}{2m}\sqrt{1-\frac{2m}{r}}\left(\sqrt{1-\frac{2m}{r}}\pm1\right)
\mathbf{s}^{\mbox{\tiny{A}}}\mathbf{s}_{\mbox{\tiny{C}}},\,
\mbox{for}\quad r>4m
\end{eqnarray}
In particular for $k=1/2$:
\begin{eqnarray}
\label{F-to}\phi_{\phantom\
\mbox{\tiny{C}}}^{\mbox{\tiny{A}}}&=&\sqrt{\sigma}\delta_{\phantom\
\mbox{\tiny{C}}}^{\mbox{\tiny{A}}}+\left(-1\pm\sqrt{2}\right)
\mathbf{s}^{\mbox{\tiny{A}}}\mathbf{s}_{\mbox{\tiny{C}}},\quad
\mbox{for}\quad \|\mathbf{s}\|^{2}_{\tilde{g}}\equiv1
\quad\mbox{or}\quad r=4m
\end{eqnarray}
Finally, we have:
\begin{eqnarray}
\label{arpa}\phi_{\phantom\
\mbox{\tiny{C}}}^{\mbox{\tiny{A}}}&=&\sqrt{\sigma}\delta_{\phantom\
\mbox{\tiny{C}}}^{\mbox{\tiny{A}}}+
\frac{-\sqrt{\sigma}\mp\sqrt{\sigma+\|\mathbf{s}\|^{2}}}{\|\mathbf{s}\|^{2}}
\mathbf{s}^{\mbox{\tiny{A}}}\mathbf{s}_{\mbox{\tiny{C}}},\quad
\mbox{for}\quad
\|\mathbf{s}\|^{2}_{\tilde{g}}\in\left]1,+\infty\right[
\end{eqnarray}
Substituting $\|\mathbf{s}\|^{2}_{\tilde{g}}$
and $\sigma$ in the (\ref{arpa})  we have:
\begin{eqnarray}
\label{organo}\phi_{\phantom\
\mbox{\tiny{C}}}^{\mbox{\tiny{A}}}&=&\delta_{\phantom\
\mbox{\tiny{C}}}^{\mbox{\tiny{A}}}-
\left(1\pm\frac{1}{\sqrt{k}}\right)\frac{1}{\left(\frac{1}{k}-1\right)}
\mathbf{s}^{\mbox{\tiny{A}}}\mathbf{s}_{\mbox{\tiny{C}}},\quad
\mbox{for}\quad k<1/2
\end{eqnarray}
or also, for $ r\in\left]2m,4m\right[$
\begin{eqnarray}
\label{vui-la}\phi_{\phantom\
\mbox{\tiny{C}}}^{\mbox{\tiny{A}}}&=&\delta_{\phantom\
\mbox{\tiny{C}}}^{\mbox{\tiny{A}}}-\frac{r}{2m}\sqrt{1-\frac{2m}{r}}
\left(\sqrt{1-\frac{2m}{r}}\pm1\right)
\mathbf{s}^{\mbox{\tiny{A}}}\mathbf{s}_{\mbox{\tiny{C}}},
\end{eqnarray}
We  note that,  the solution (\ref{arpa}) matches
(\ref{pan-for-1}).

Alternatively it is possible to deform  the metric (\ref{Per-dri}) in a flat
metric  by   changing the coordinate $r$
into the coordinate $R$ defined as:
\begin{equation}\label{Sa11}
R\equiv\frac{1}{2}\left(r\sqrt{k}+r-m\right), \quad
\mbox{where}\quad r=R\left(1+\frac{m}{2R}\right)^{2}.
\end{equation}
We can write $\tilde{g}$ as
\begin{equation}\label{250}
\tilde{g}=\omega^{\hat{r}}\otimes
\omega^{\hat{r}}+\omega^{\hat{\theta}}\otimes
\omega^{\hat{\theta}}+\omega^{\hat{\phi}}\otimes
\omega^{\hat{\phi}},
\end{equation}
in the  triad $\omega^{\mbox{\tiny{A}}}_{a}$ defined as
\begin{equation}\label{collo}
\omega^{\hat{r}}=\left(1-\frac{m^2}{4R^{2}}\right)^{-1}dr,\quad
\omega^{\hat{\theta}}=R d\theta,\quad\omega^{\hat{\phi}}=R
\sin\theta d\phi
\end{equation}
%%as
while  the metric $\hat{g}$ is simplify a conformal deformation of
$\tilde{g}$
\begin{equation}\label{Cam-Bio}
\hat{g}=\sigma\tilde{g},\quad
%\end{equation}
%%
%where now
%%
%\begin{equation}\label{Ottobre 2008}
\sigma\equiv\left(1+\frac{m}{2R}\right)^{4}
\end{equation}
and $\mathbf{s}$ is the zero--vector. In this case the
deforming matrix is
\begin{equation}\label{Un-cil-cap}
\phi_{\phantom\
\mbox{\tiny{C}}}^{\mbox{\tiny{A}}}=\sqrt{\sigma}\Lambda_{\phantom\
\mbox{\tiny{C}}}^{\mbox{\tiny{A}}}.
\end{equation}
\subsection{Three--dimensional Kerr space}\label{Sec:Kerr}
Consider the following Kerr metric in the  Boyer--Lindquist
coordinates:
%
%\begin{equation}\label{USA}
%g=-\frac{\Delta}{\rho^2}\left[dt-a\sin^2\theta
%d\phi\right]^{2}+\frac{\sin^2\theta}{\rho^2}\left[a dt-\left(r^2+a^2\right)d\phi\right]^{2}+\frac{\rho^2}{\Delta}dr^2+\rho^2d\theta^2,
%\end{equation}
\be\label{USA}
g=-dt^2+\frac{\rho^2}{\Delta}dr^2+\rho^2
d\theta^2+(r^2+a^2)\sin^2\theta
d\phi^2+\frac{2m}{\rho^2}r(dt-a\sin^2\theta d\phi)^2\ ,
\ee
where
\begin{equation}\label{Pre-ta}
\rho^2\equiv r^2+a^2\cos\theta^2, \quad \Delta\equiv r^2-2mr+a^2.
\end{equation}
and $\rho^2-2mr>0$; here $m$ is a mass parameter and the specific angular momentum is given as $a=J/m$, where $J$ is the
total angular momentum of the gravitational source,  we
consider the Kerr  black hole spacetime  defined by $a\in ]0, m[ $, the extreme black hole source is for  $a=m$, the  limiting case $a=0$ is the Schwarzschild solution \cite{wald,Kramer-ste-M,Banados:1992wn,Banados:1992gq,BainGabac28,Cho08,GrifPod09}.

We consider the  stationary metric $\hat{g}$ defined as
\begin{equation}\label{lA-prime}
\hat{g}\equiv\frac{\rho^2}{r^2}\left(\frac{r^2}{\Delta}dr\otimes
dr+r^2d\theta\otimes d\theta+\frac{\mathcal{A}}{\rho^4 }r^2 \sin\theta^2
d\phi\otimes d\phi\right),
\end{equation}
where $\mathcal{A}\equiv\left(r^2+a^2\right)^2-\Delta  a^2\sin\theta^2$.
 First  we  change the coordinate $r$ into the
coordinate $R\equiv R_+$   defined as follows
\begin{equation}\label{Sal1Ot-08}
R_{\pm}\equiv\frac{1}{2}\left(r-m\pm\sqrt{\Delta}\right), \quad
\mbox{where}\quad
r=R\left[\left(1+\frac{m}{2R}\right)^{2}-\frac{a^2}{4R^{2}}\right].
\end{equation}
for $a=0$ this reduces to the isotropic radius for the Schwarzschild metric.
We can introduce the metric tensor  $\tilde{g}$
\begin{equation}\label{250bis}
\tilde{g}=\omega^{\hat{r}}\otimes
\omega^{\hat{r}}+\omega^{\hat{\theta}}\otimes
\omega^{\hat{\theta}}+\omega^{\hat{\phi}}\otimes
\omega^{\hat{\phi}},
\end{equation}
in the   triad $\omega^{\mbox{\tiny{A}}}_{a}$, defined as
\begin{equation}\label{coll-dol}
\omega^{\hat{r}}=\left(1-\frac{m^2}{4R^{2}}+\frac{a^2}{4R^2}\right)^{-1}dr,\quad
\omega^{\hat{\theta}}=R d\theta,\quad\omega^{\hat{\phi}}=\sqrt{\frac{\mathcal{A} R^2}{\rho^4}-\frac{a^2m^2 }{(\rho^2-2 m r)} } \sin\theta d\phi
\end{equation}
%%as
while  the metric $\hat{g}$ reads
\begin{equation}\label{Sal-Ma}
\hat{g}=\sigma \tilde{g}+\textbf{s}\otimes\textbf{s}
\end{equation}
where $\epsilon=+1$ and
\begin{equation}\label{allNu}
\sigma\equiv\frac{\rho^2}{R^{2}}, \quad \mathbf{s}\equiv\frac{\rho a\sin
\theta}{R\sqrt{\rho^2-2mr}}\omega^{\hat{\phi}},
\end{equation}
and
\begin{equation}\label{serioSul}
\|\mathbf{s}\|^{2}_{\tilde{g}}=\sigma\left(\frac{a^2\sin^2
\theta}{\rho^2-2mr}\right)>0.
\end{equation}
 The  deforming matrix $\phi_{\phantom\
\mbox{\tiny{C}}}^{\mbox{\tiny{A}}}$ associated to the deformation
(\ref{Sal-Ma}) has the form (\ref{per-ta}). In particular
Eq.(\ref{eq}) has the following solutions:
\begin{eqnarray}
\label{PA-FOR}\phi_{\phantom\
\mbox{\tiny{C}}}^{\mbox{\tiny{A}}}&=&\sqrt{\sigma}\delta_{\phantom\
\mbox{\tiny{C}}}^{\mbox{\tiny{A}}}+\frac{-\sqrt{\sigma}\mp
\sqrt{\sigma+\|\mathbf{s}\|^{2}}}{\|\mathbf{s}\|^{2}}
\mathbf{s}^{\mbox{\tiny{A}}}\mathbf{s}_{\mbox{\tiny{C}}},\quad
\mbox{for}\quad
\|\mathbf{s}\|^{2}_{\tilde{g}}\in\left]0,\sigma\right[
\end{eqnarray}
and substituting in the (\ref{PA-FOR}) the explicit forms (\ref{allNu}) and (\ref{serioSul}) for
$\|\mathbf{s}\|^{2}_{\tilde{g}}$ and $\sigma$   respectively we obtain:
\begin{eqnarray}
\label{PA-FOR2}\phi_{\phantom\
\mbox{\tiny{C}}}^{\mbox{\tiny{A}}}&=&\frac{\rho}{R}\delta_{\phantom\
\mbox{\tiny{C}}}^{\mbox{\tiny{A}}}-\frac{R}{\rho}\frac{\sqrt{\rho^2-2mr}}{a^2\sin^2\theta}
\left(\sqrt{\rho^2-2mr}\pm\sqrt{\Delta}
\right)\mathbf{s}^{\mbox{\tiny{A}}}\mathbf{s}_{\mbox{\tiny{C}}},\,
\end{eqnarray}
for
\begin{eqnarray}
\label{PA-FOR3} \rho^2-2mr>  a^2 \sin^2\theta.
\end{eqnarray}
In particular:
\begin{eqnarray}\label{tamb-ro}
\phi_{\phantom\ \mbox{\tiny{C}}}^{\mbox{\tiny{A}}}
&=&\frac{\rho}{R}\delta_{\phantom\
\mbox{\tiny{C}}}^{\mbox{\tiny{A}}}+\frac{R}{2\rho}
\mathbf{s}^{\mbox{\tiny{A}}}\mathbf{s}_{\mbox{\tiny{C}}},\,
\mbox{for}\quad \|\mathbf{s}\|^{2}_{\tilde{g}}\equiv0 \quad\mbox{or
 also}\quad a\sin\theta=0
\end{eqnarray}
meanwhile:
\begin{eqnarray}
\nonumber\phi_{\phantom\
\mbox{\tiny{C}}}^{\mbox{\tiny{A}}}&=&\frac{\rho}{R}\delta_{\phantom\
\mbox{\tiny{C}}}^{\mbox{\tiny{A}}}-\frac{R}{\rho}\left(1\pm\sqrt{2}\right)
\mathbf{s}^{\mbox{\tiny{A}}}\mathbf{s}_{\mbox{\tiny{C}}},\\
\label{tri-ang} \mbox{for}&\quad&
\|\mathbf{s}\|^{2}_{\tilde{g}}\equiv\sigma \quad\mbox{or also}\quad
\rho^2-2mr= a^2 \sin^2\theta.
\end{eqnarray}
 Finally, we have:
\begin{eqnarray}
\label{p-tti}\phi_{\phantom\
\mbox{\tiny{C}}}^{\mbox{\tiny{A}}}&=&\sqrt{\sigma}\delta_{\phantom\
\mbox{\tiny{C}}}^{\mbox{\tiny{A}}}+\frac{-\sqrt{\sigma}\mp
\sqrt{\sigma+\|\mathbf{s}\|^{2}}}{\|\mathbf{s}\|^{2}}
\mathbf{s}^{\mbox{\tiny{A}}}\mathbf{s}_{\mbox{\tiny{C}}},\,
\mbox{for}\quad
\|\mathbf{s}\|^{2}_{\tilde{g}}\in\left]\sigma,+\infty\right[
\end{eqnarray}
and using Eqs.\il(\ref{allNu}) and (\ref{serioSul}) in the (\ref{p-tti}) we have:
\begin{eqnarray}
\nonumber\phi_{\phantom\
\mbox{\tiny{C}}}^{\mbox{\tiny{A}}}&=&\frac{\rho}{R}\delta_{\phantom\
\mbox{\tiny{C}}}^{\mbox{\tiny{A}}}-\frac{R}{\rho}
\frac{\sqrt{\rho^2-2mr}}{a^2\sin^2\theta}
\left(\sqrt{\rho^2-2mr}\pm\sqrt{\Delta}\right)
\mathbf{s}^{\mbox{\tiny{A}}}\mathbf{s}_{\mbox{\tiny{C}}},\\\nonumber\\
\label{PA-FOR26}&&\mbox{for}\quad 0<\rho^2-2mr< a^2 \sin^2\theta .
\end{eqnarray}
We note that the solution (\ref{p-tti}) matches
(\ref{PA-FOR}).
\section{Discussion and conclusions}\label{Sec:conclusion-disc}
In this work we studied deformations of three dimensional  metrics. We related  two different procedures to obtain a deformed  metric tensor of a three dimensional manifold.   We started considering the procedure outlined in \cite{LlosaeSoler}  where a deformation of  three-dimensional metric tensor $h$ was obtained by first multiplying it by a conformal transformation with  a  scalar field $ \sigma$  and then summing it with   the tensor product $\mathbf{s} \otimes \mathbf{s}$   of the  differential 1--form
  $\mathbf{s}$.  The deformation so defined is be applied to  attain  a  constant curvature Riemannian metric.  However, Theorem \ref{th1}
is proved  only locally, in the neighborhood of the Cauchy surface,  and in the analytical case where all the data  are real analytic functions.
Then we introduced the metric deformations according to \cite{Cap-S07}. They  seem to be more general because they
allow  to define a transformation between any two metrics by means of generic, not necessarily real, scalar fields.
%This deformation
%can be applied in any $n$-dimensional  manifold $M$, through the  general choice of the  vielbeins   and the scalar fields
%defined on $M$.
%Thus, here  we restricted the analysis to the case of constant curvature
%metric and
By writing the  algebraic equations which identified the  metric transformations of  \cite{LlosaeSoler}  with the single components of the deforming matrices, we found the explicit expressions  for the generic  deforming matrices.  We discussed separately  the real and complex solutions and provided  a classification of various deforming matrices according to space-time properties of $ \sigma$  and of  $\mathbf{s}$.  Since the deformations obtained are not  just  conformal transformations, we expect  that     the casual structure  of the  deformed manifold  is modified and we found the conditions under which it is preserved.

Finally, as an application, we presented   in terms of a  flat metric, to the geometry  of a the three--dimensional sphere  and  of   the spatial sector of a Kerr space-time.

While the work of Coll et al. gives an interesting approach to deforming metrics it seems to be an extension of the family of  the Kerr-Schild metrics. The deformation by scalar matrices seems to be more general. And one of its  main advantages  is that it is covariant with respect to coordinate transformations, the price payed for this is that the deforming matrices are not uniquely defined.

We expect deformations  to be amenable to different physical  applications in general relativity and extended theories of gravity.  For example they can be applied to cast the solutions of the gravitational equations in a suitable manner according to astrophysical observations or for studying numerical implementations.   Furthermore  it is possible to use the space-time metric deformations  as geometric perturbations.

These aspects  of space-time deformations  and other  considerations on  the  possible associated phenomenology,  as gravitational lensing and redshift, have been addressed  in \cite{Tesi,Pugliese:2009we},  where it is  discussed in particular the possibility to  constraint the deforming scalar fields.  In this respect, these considerations can be
directions this work could take in the future from the results in Section (\ref{Sec:three-metric}) and to be investigated elsewhere.

The present analysis has been restricted to the case  of three dimensional metrics, but a more  general version of these results for arbitrary dimensions  is also possible.

\section*{Acknowledgments}

DP gratefully acknowledges support from the Blanceflor
Boncompagni-Ludovisi, n\'ee Bildt   and wishes to thank  the Angelo
Della Riccia Foundation and thanks the institutional support
of   the Faculty of Philosophy and Science of the Silesian University of Opava.
This work is partially supported by the INFN Iniziativa Specifica QGSKY ``Quantum Universe".
%\appendix
%

%\addcontentsline{toc}{chapter}{Bibliografia}


\begin{thebibliography}{99}

\bibitem{three--dim}
Coll B.,   Llosa J., and  Soler D. , {Three-dimensional metrics
as deformations of a constant curvature metric},
Gen.\ Rel.\ Grav.\ {\bf 34},  269 (2002).

%\cite{Ellis:1987zz}
\bibitem{Ellis:1987zz}
  Ellis G.~F.~R.~ and Stoeger W.~,
  The 'fitting problem' in cosmology,
  Class.\ Quant.\ Grav.\  {\bf 4}, 1697  (1987).
 


\bibitem{AcenaKroon11}  Acena  A. \& Valiente Kroon J. A. , { Conformal extensions for stationary spacetimes}, Class. Quantum Grav. 28, 225023 (2011).

%%
   \bibitem{Frie02G}
Friedrich H., { Conformal Einstein evolution, in The conformal structure of spacetime: Geometry, Analysis, Numerics}, edited by J. Frauendiener \& H. Friedrich, Lecture Notes in
Physics,  Springer (2002).

\bibitem{LuKroon12G}
 L\"ubbe C. \&  Valiente Kroon J. A.,
{ The extended Conformal Einstein field equations with
matter: the Einstein-Maxwell system},
 J. Geom. Phys. \textbf{62}, 1548 (2012).

\bibitem{LuKroon13An}
  L\"ubbe C. and Valiente Kroon J.~A.,
  A Class of Conformal Curves in the Reissner-Nordstr\"om Spacetime,
  Annales Henri Poincare {\bf 15}, 1327 (2014).

\bibitem{LuKroon13}L\"ubbe C.~ and Kroon J.~A.~V.~,
  A conformal approach for the analysis of the non-linear stability of pure radiation cosmologies,
  Annals Phys.\  {\bf 328},  1 (2013).
 % [arXiv:1111.4691 [gr-qc]].
  %%CITATION = ARXIV:1111.4691;%%
  %10 citations counted in INSPIRE as of 15 Oct 2014
\bibitem{CFEBook}
 Valiente Kroon J. A.,    ``Conformal methods in General Relativity'' Cambridge University Press, in preparation.



\bibitem{Capozziello:2009dz}
  Capozziello S.~ and De Laurentis M.,
  A Review about Invariance Induced Gravity: Gravity and Spin from Local Conformal-Affine Symmetry,
  Found.\ Phys.\  {\bf 40}, 867 (2010).



\bibitem{Capozziello:2011et}
  Capozziello S. and De Laurentis M.,
  Extended Theories of Gravity,
  Phys.\ Rept.\  {\bf 509},  167 (2011).

 \bibitem{TerKazarian:2011kh}
  Ter-Kazarian G.,
  Two-step spacetime deformation induced dynamical torsion,
  Class.\ Quant.\ Grav.\  {\bf 28}, 055003 (2011).




  \bibitem{carfandmarz}
Carfora M.   and Marzuoli   A., Smoothing Out Spatially Closed Cosmologies, Phys. Rev. Lett. {\bf 53},  2445 (1984).
 %

  \bibitem{Mukhanov:1990me}
  Mukhanov V.~F.~, Feldman H.~A.~ and Brandenberger R.~H.~,
  Theory of cosmological perturbations. Part 1. Classical perturbations. Part 2. Quantum theory of perturbations. Part 3. Extensions,
  Phys.\ Rept.\  {\bf 215} 203,  (1992).
  %%CITATION = PRPLC,215,203;%%
  %2002 citations counted in INSPIRE as of 14 Oct 2014
%\cite{Newman:1965tw}
\bibitem{Newman:1965tw}
 Newman  E.~T.~ and Janis A.~I.,
  Note on the Kerr spinning particle metric,
  J.\ Math.\ Phys.\  {\bf 6} 915,  (1965).


  %\cite{Carlip:1994gc}
\bibitem{Carlip:1994gc}
  Carlip S.~ and Teitelboim C.~T,
  Aspects of black hole quantum mechanics and thermodynamics in (2+1)-dimensions,
  Phys.\ Rev.\ D {\bf 51},  622(1995).
  %[gr-qc/9405070].
  %%CITATION = GR-QC/9405070;%%
  %123 citations counted in INSPIRE as of 14 Oct 2014


\bibitem{LlosaeSoler}
  Llosa J.~ and ~Soler D.,
  {On the degrees of freedom of a semi-Riemannian metric},
  Class.\ Quant.\ Grav.\  {\bf 22}, 893 (2005).

  %6 citations counted in INSPIRE as of 15 Oct 2014
\bibitem{Soler:2005xa}
  Soler D.,
  Reference frames and rigid motions in relativity: Applications,
  Found.\ Phys.\  {\bf 36}, 1718  (2006).

  \bibitem{Llosa:2008yk}
 Llosa  J.~ and Carot  J.,
  Flat deformation theorem and symmetries in spacetime,
  Class.\ Quant.\ Grav.\  {\bf 26}, 055013 (2009).
  %%CITATION = ARXIV:0809.1030;%%
  %4 citations counted in INSPIRE as of 15 Oct 2014

\bibitem{Llosa:2009sc}
  Llosa J.~ and Carot  J.,
  Flat deformation of a spacetime admitting two commuting Killing fields,
  Class.\ Quant.\ Grav.\  {\bf 27}, 245006  (2010).


\bibitem{Coll:2000rm}
Coll B., Hildebrandt S.~R. and ~Senovilla J.~M.~M.,
  Kerr-Schild symmetries,
  Gen.\ Rel.\ Grav.\  {\bf 33},  649  (2001).

  \bibitem{Coll1} B. Coll, {A universal law
of gravitational deformation for general relativity}, Proc. of the
Spanish Relativistic Meeting, EREs, Salamanca Spain (1998).
\bibitem{Cap-S07}
 Capozziello S.~ and Stornaiolo C.~,
  Space-time deformations as extended conformal transformations,
  Int.\ J.\ Geom.\ Meth.\ Mod.\ Phys.\  {\bf 5} 185, (2008).

\bibitem{Tesi}
  Pugliese D., { Deformazioni di metriche spaziotemporali}, Thesis
(unpublished) 2006--2007.   `` Universit\`{a} degli studi di Napoli Federico II''.
Dipartimento di Scienze Fisiche, Biblioteca \emph{Roberto
Stroffolini}.
%
%
\bibitem{Llosa:2003di}
Llosa J.~ and Soler D.~, {Reference frames and rigid motions
in relativity}, Class.\ Quant.\ Grav.\  {\bf 21}, 3067  (2004).

%\cite{Pugliese:2009we}
\bibitem{Pugliese:2009we}
  Pugliese D.~, Stornaiolo C.~ and CapozzielloS.,
Deformations of spacetime metrics,
  arXiv:0910.5738 [gr-qc].
  %%CITATION = ARXIV:0910.5738;%%

%\bibitem{4--dim} B. Coll
%\emph{A Universal Law of Gravitational Deformation for General
%Re\-la\-ti\-vi\-ty} in Proceedings oh the Spanish Relativity meeting
%in honour of 65th birthday of Luis Bell ""Gravitation and Relativity
%in General'' ed by Martin, J. Ruiz, E. Atrio, F. Molina, A. World
%Scientific.(1999)
%%
%\bibitem{quadridimensionale} B. Coll
%\emph{About deformation and rigidity in relativity}  <http:\
%//Syrte.obspm.fr\ /$\sim $Coll/>   agg.2005
%
%\bibitem{three--dim} B. Coll, J. Llosa, D. Soler
%{Three--Dimensional Metrics as Deformations of a Constant
%Curvature Metric} Vol.34, No.2, (2002)
%%
%\bibitem{LlosaeSoler}
%J. Llosa and D. Soler \emph{On the degrees of freedom of a
%semi-Riemannian metric} Class. Quant. Grav. 22, 893 (2005)
%



\bibitem{wald}  Wald R. M. , {General Relativity}, The University of
Chicago Press (1984).

\bibitem{Kramer-ste-M}
Stephani H.,  Kramer D., MacCallum  M.,
Exact Solutions of Einstein's Field Equations, Cambridge Monographs on Mathematical Physics (2009).





\bibitem{Banados:1992wn}
  Banados M.~, Teitelboim  C.~ and Zanelli J.~,
   {The Black hole in three-dimensional space-time},
  Phys.\ Rev.\ Lett.\  {\bf 69}, 1849 (1992).

\bibitem{Banados:1992gq}
 Banados M., Henneaux M., Teitelboim C. and Zanelli J.,
   {Geometry of the (2+1) black hole},
  Phys.\ Rev.\  D {\bf 48} 1506, (1993)






\bibitem{BainGabac28} Dain S. \&  Gabach-Clement M. E.,
{ Small deformations of extreme Kerr black hole initial
data}, Class. Quantum Grav. 28, 075003 (2011)




\bibitem{Cho08}
Y.~Choquet-Bruhat,
 { General Relativity and the Einstein equations},
 Oxford University Press (2008).

\bibitem{GrifPod09}
Griffiths J. B., Podolský J.,
Exact Space-Times in Einstein's General Relativity, Cambridge Monographs on Mathematical Physics (2009).



%\bibitem{AJSHamilton}
%Andrew J. S. Hamilton "General Relativity, Black Holes, and Cosmology" available only in
%\url{http://www.texample.net/tikz/resources/}http://jila.colorado.edu/ajsh/phys5770 14/notes.html}
%
\end{thebibliography}
\end{document}